\documentclass[useAMS,usenatbib]{mn2e}

\input{epsf}
\newcommand{\mic}{\,$\mu$m}
\newcommand{\kms}{\,km s$^{-1}$}
\newcommand{\dg}{$^{\circ}$}
\newcommand{\Wma}{\,W~m$^{-2}$~arcsec$^{-2}$}
\newcommand{\Msolar}{\,M$_{\odot}$}

\newcommand{\htwo}{H$_2$\,2.122\,$\mu$m}

\title[Observations of jets in DR21/W75]
{WFCAM, Spitzer-IRAC and SCUBA observations of the massive star forming
region DR21/W75: I. The collimated molecular jets}
\author[C.J.~Davis et al.]
       {C.J.~Davis$^{1}$,
        M.S.N.~Kumar$^{2}$,
        G.~Sandell$^{3}$, 
        D.~Froebrich$^{4}$,
        M.D.~Smith$^{5}$ and M.J.\newauthor
        Currie$^{6}$ \\
$^1$Joint Astronomy Centre, 660 North A'oh\={o}k\={u} Place,
        University Park, Hilo, Hawaii 96720, USA. \\
$^2$Centro de Astrofisica da Universidade do Porto,
    Rua das Estrelas s/n 4150-762 Porto, Portugal \\
$^3$SOFIA-USRA, NASA Ames Research Center, MS N211-3, Moffett Field, CA 94035 \\
$^4$Dublin Institute for Advanced Studies, 5 Merrion Square, Dublin 2, Ireland \\
$^5$Centre for Astrophysics \& Planetary Science,
    School of Physical Sciences, The University of Kent,
    Canterbury CT2 7NR, England \\
$^6$Rutherford Appleton Laboratory, Didcot, Oxfordshire, OX11 0QX, England
}

\begin{document}

\date{Accepted 2006 ... ; 
      Received 2006 ... ; 
      in original form 2006 August 31}

\pagerange{\pageref{firstpage}--\pageref{lastpage}} \pubyear{2006}

\maketitle

\label{firstpage}

\begin{abstract} 

We present wide-field near-infrared images of the DR21/W75 high-mass
star forming region, obtained with the Wide Field Camera, WFCAM, on
the United Kingdom Infrared Telescope.  Broad-band JHK and narrow-band
H$_2$ 1-0S(1) images are compared to archival mid-IR images from the
Spitzer Space Telescope, and 850\micron\ dust-continuum maps obtained
with the Submillimeter Common User Bolometer Array (SCUBA). Together
these data give a complete picture of dynamic star formation across
this extensive region, which includes at least four separate star
forming sites in various stages of evolution.  The H$_2$ data reveal
knots and bow shocks associated with more than 50 individual flows.
Most are well collimated, and at least five qualify as parsec-scale
flows.  Most appear to be driven by embedded, low-mass protostars.
The orientations of the outflows, particularly from the few
higher-mass sources in the region (DR21, DR21(OH), W75N and ERO~1),
show some degree of order, being preferentially orientated roughly
orthogonal to the chain of dusty cores that runs north-south through
DR21.  Clustering may inhibit disk accretion and therefore the
production of outflows; we certainly do not see enhanced outflow
activity from clusters of protostars.  Finally, although the low-mass
protostellar outflows are abundant and widely distributed, the current
generation does not provide sufficient momentum and kinetic energy to
account for the observed turbulent motions in the DR21/W75 giant
molecular clouds.  Rather, multiple epochs of outflow activity are
required over the million-year timescale for turbulent decay.

\end{abstract}




\section{Introduction}

Star formation studies have in recent years benefited from a fresh,
``large scale'' view, via observations of Giant Molecular Clouds
(GMCs) and large clusters of clumps and cores at infrared and
(sub)millimetre wavelengths, and through studies of the interplay
between the thousands of evolving young sources within each region.

Infrared survey instrumentation that offers an unparalleled field
of view at high spatial resolution is now, or will soon be, available
at modest-sized telescopes across the globe.  One such instrument is
the near-infrared (near-IR) Wide Field Camera, WFCAM, at the
U.K. Infrared Telescope (UKIRT).  WFCAM provides deep imaging of
degree-sized regions in broad and narrow-band filters in just a few
hours.  When combined with photometric data at longer wavelengths,
e.g. from the Submillimeter Common User Bolometer Array (SCUBA - soon
to be replaced by SCUBA-2) and the Spitzer Space Telescope, obtaining
a complete picture of evolving star formation (stellar populations,
the IMF, the clump mass function, etc.) becomes
possible. Complementary wide-field optical images of Herbig-Haro (HH)
jets, and future large-scale (sub)millimetre surveys of molecular
``CO'' outflows, may also be used to examine the dynamic processes
that shape -- through the injection of turbulence or triggered cloud
collapse -- star formation across each GMC.


One region that has already benefited from wide-field observations
with Spitzer, SCUBA and now WFCAM is the DR21/W75S,
DR21(OH) and W75N complex of massive star forming cores in Cygnus X
\citep[][for a more recent overview of the Cygnus X complex of GMCs 
see Schneider et al. 2006]{wes58,dow66}.  The DR21 core contains two
well-studied cometary H{\sc ii} regions \citep{har73,dic86,cyg03} and
an H$_2$O maser \citep{gen77}, while DR21(OH), a dense core roughly
3\arcmin\ to the north of DR21, is abundant with methanol, water and
OH masers \citep[][and references therein]{lie97}.  DR21(OH) is
undetected in radio continuum emission and so is probably younger than
DR21.  W75N, roughly 20\arcmin\ to the north of DR21, comprises a
group of embedded infrared sources \citep{moo91b,per06}, H{\sc ii}
regions and compact radio sources
\citep[e.g.][]{has81,hun94}, as well as H$_2$O and OH masers
\citep{hun94,tor97}.  


Early work by \citet{dic78} and \citet{fis85} showed that DR21 and W75N
coincide with two, possibly interacting molecular cores along a north-south
molecular ridge extending over at least half a degree.  These cores,
evident in the more recent 850\mic\ dust continuum maps of \citet{val06} and
\citet{gib07}, have been resolved into smaller clumps by various
groups \citep{wil90,man92,cha93a,she01}. Spitzer images at 5.8 and
8.0\mic\ reveal streamers of emitting material that seem to radiate
away from the brighter infrared sources located along the molecular
ridge \citep{mar04}, features possibly produced by ionization and ablation
of the less dense material on either side of the ridge.
\citet{mar04} also identify a number of ``extremely red objects'', or
EROs, along the ridge.  These are luminous, highly-reddened objects
that may be very young, massive, early B-type sources.  


In terms of outflows, DR21 and W75N have been studied independently by
a number of groups.  DR21 drives one of the most spectacular flows
known \citep{gar91b,gar92}.  The bipolar outflow is extremely bright
in H$_2$ line emission \citep{gar91a,dav96,smi98} and produces
methanol abundance enhancements and masers through shocks
\citep{lie97}.  The dominant outflow from W75N is somewhat fainter in H$_2$,
although it appears to be a good example of a wind-driven flow, in
which sweeping bow shocks entrain ambient gas to form a massive
molecular outflow \citep{dav98a,dav98b}.  Early CO maps showing the
orientation and bipolarity of the W75N outflow were presented by
\citet{fis85}; more detailed maps in CO and millimetre continuum
emission by \citet{she01}, \citet{she03} and \citet{she04} illustrate
the complexity of the region.


In this paper we discuss WFCAM data obtained in H$_2$ 1-0S(1) emission
and in the J, H and K bands.  The images cover a 0.8\degr $\times$
0.8\degr\ ``tile'' (see Section 2) centred close to DR21.  The
broad-band photometry are discussed in detail in a companion paper
\citep[][hereafter Paper II]{kum06}; in this work we focus on the H$_2$ 
emission-line features associated with the many outflows along the
DR21/W75N ridge.  A comparison is also made with SCUBA 850\mic\ data
(see also \citet{val06} and \citet{gib07} for a similar map), and
archival Spitzer IRAC images at 3.6\mic , 4.5\mic , 5.8\mic\ and
8.0\mic\ \citep[originally presented by][]{mar04}.  The combined
SPITZER and SCUBA maps suggest that stars are forming within a
flattened north-south filament viewed almost edge on.  One of the
goals of the WFCAM observations was to search for outflows ejected
east-west, i.e. perpendicular to this dense filament.  We discuss the
effect these outflows might have on the general GMC, and look for
examples of collimated jets from the more massive YSOs.

There has been some disagreement on the distance to DR21 and W75N,
although both are thought to be in the range 1.5-3.0~kpc
\citep{gen77,fis85,ode93,sch06}.  In Paper II we find that the upper
limit gives a more realistic population of massive stars, so we adopt
3~kpc as a general distance in this paper.


\section{Observations}

\subsection{WFCAM and Spitzer imaging}

Wide-field images of the DR21/W75 region were obtained during
service observing at the United Kingdom Infrared Telescope (UKIRT) on
6 May, 11 May and 9 June 2005, using the near-IR wide-field camera
WFCAM \citep{cas06}. The camera employs four Rockwell Hawaii-II
(HgCdTe 2048x2048) arrays spaced by 94\% in the focal plane.  The
pixel scale measures 0.40\arcsec . 

To fill in the gaps between the arrays, and thereby observe a
contiguous square field on the sky covering 0.75 square degrees --
what is referred to as a WFCAM ``tile'' -- observations at four
positions are required.  At each position, to correct for bad pixels
and array artifacts, a five-point jitter pattern was executed (with
offsets of 3.2\arcsec\ or 6.4\arcsec ); to fully sample the seeing, at
each jitter position, a 2x2 micro-stepped pattern was also used, with
the array shifted by half a pixel.  In this way 20 frames were
obtained at each of the four positions in the tile.

Data through {\em Mauna Kea Consortium} broad-band J, H and K filters,
and a narrow-band H$_2$ 1-0S(1) filter ($\lambda=2.121$\mic ,
$\delta\lambda=0.021$\mic ), were obtained \citep{hew06}. Exposure
times of 5\,sec and 40\,sec were used with the broad and narrow-band
filters, respectively.  A correlated double sampling (CDS) readout
mode was used with the broad-band filters, while a non-destructive
readout (NDR) mode was employed with the longer-exposure H$_2$ images.
The latter yields a lower read noise ($\sim$20e- as compared to
$\sim$30e- with CDS) at the expense of greater amplifier glow and a
higher risk of electronic pickup. With the 5-point jitter pattern and
2x2 micro-stepping in each broad-band filter, the total
on-source/per-pixel integration time was 100~sec per filter.  In H$_2$
the whole 20-frame sequence was repeated; hence, the total per-pixel
exposure time was 1600\,sec.

The data were reduced by the Cambridge Astronomical Survey Unit
(CASU), which is a fore-runner to the VISTA Data Flow System Project
(VDFS), a collaborative effort between the Universities of London,
Cambridge and Edinburgh in the U.K. CASU is responsible for the design
and implementation of the data processing pipeline used prior to the
archiving and subsequent distribution of all WFCAM data.  The
reduction steps are described in detail by \citet{dye06} and
\citet{irw06}.  Briefly, dark frames secured at the beginning of the
night were subtracted from each target frame to remove bad
pixels, amplifier glow, and reset anomalies from the raw frames.
Twilight flats were then used to flat-field the data.  
The micro-stepped frames were interleaved and the offset frames
combined by weighted averaging (using a confidence map derived largely
from the flat-field frames) to produce ``leavestack'' frames.

Astrometric calibration was achieved through a cubic radial fit and a
six-parameter linear transformation to objects found in the 2MASS point
source catalogue \citep{dye06}.  The transformation takes into account
translation, scaling, rotation and shear. Photometric
calibration was also achieved through the use of 2MASS data
\citep{dye06,hew06}.

All reduced WFCAM data are archived at the WFCAM Science Archive
(WSA), which is part of the Wide Field Astronomy Unit (WFAU) hosted by
the Institute for Astronomy at the Royal Observatory, Edinburgh,
U.K. (http://surveys.roe.ac.uk/wsa/index.html). WSA data are
distributed as RICE-compressed Multi-Extension Fits (MEF) files. The
data for one filter covering the 0.8\dg $\times$0.8\dg\ tile appear as
four leavestack files in the database, one MEF file for each of the
four offsets on sky needed to cover the tile.  Each leavestack file
therefore contains four individual images, one for each of the four
WFCAM cameras.

Having retrieved the reduced data from the WFCAM Science Archive
(WSA), Starlink software were used to remove residual large-scale
gradients across each of the 16 images in each tile.  The
Starlink:KAPPA command SURFIT was used to fit a coarse surface to the
background in each frame; the four surface fit images in each corner
of the full tile were median-averaged and used to flat-field the
four associated target frames.  The frames were subsequently mosaicked
together using the KAPPA:WCSALIGN routine applied to the world
coordinates assigned to each frame by the CASU pipeline, together with
the Starlink:CCDPACK mosaicking routine MAKEMOS. These gave one
complete tile image in each filter.

The Spitzer Space Telescope images discussed in this paper were
retrieved from the Spitzer archive.  Our processing of these data is
described in Paper II.  The observations were obtained with the IRAC
camera, which is described by \citet{faz04}. The data also appear in
recent papers by \citet{mar04} and \citet{per06},
although \citet{mar04} only give preliminary results, while
\citet{per06} focus only on a small $\sim$2\arcmin $\times$2\arcmin\
field centred on W75N.  

\subsection{SCUBA observations}

The sub-millimetre data were obtained with the SCUBA bolometer array camera
\citep{hol99} on the James Clerk Maxwell Telescope (JCMT) on Mauna Kea,
Hawaii on four partial nights from May 26 to May 29, 2001 in marginal
sub-millimetre conditions; the precipitable water vapor (PWV) was
$\sim$1.3--2~mm (optical depth at 850\mic\ $\sim$0.25--0.36), although
it was fairly stable each night. Due to the high PWV content in the
atmosphere only the 850\mic\ data were usable. The standard
``scan-mapping'' mode \citep{jen98} was employed with six different
chop throws: 30\arcsec, 44\arcsec, and 68\arcsec\ in Right Ascension
and Declination respectively, while scanning in the Nasmyth coordinate
frame.  Each chop throw was repeated 2--4 times depending on airmass
and sky opacity, the goal being to reach the same noise level in each
dual beam map.  The rms noise level in the final co-added dual beam
maps was $\sim$0.09--0.1 Jy~beam$^{-1}$. The pointing was checked
frequently on the point source MWC~349, and each dual beam map was
corrected for any pointing drift between pointing observations. Based
on the pointing data the positional accuracy in the final 850\mic\
image is estimated to be $\leq$1\arcsec. The maps were calibrated with
scan maps of Uranus as the primary flux calibrator and CRL~2688 as the
secondary calibrator. The flux calibration is estimated to be good to
within 5\% with respect to Uranus. The telescope half power beam width
was measured to be 15.2\arcsec\ from scan maps of Uranus.

The maps were reduced in a standard way using the Starlink reduction
packages SURF and Kappa \citep{san01}, except that the individual dual
beam maps where created by weighting the noise in each individual
bolometer using the SURF task {\em setbolwt}, which reduced the noise
level in the maps by $\sim$5--10\% . We then estimated the rms noise
level using emission free areas in the maps and co-added the maps by
accounting for the difference in noise between individual maps for the
same chop throw and chop position angle. The final calibrated 850\mic\
map has a noise level of 50--60 ~Jy~beam$^{-1}$ over the whole
area. For further analysis we divided the map into three areas (south,
middle, and north), removed the error beam using the MIRIAD task CLEAN
\citep{sau95} and restored the maps to 14.0\arcsec\ resolution.  Below
we present only photometry of the major 850\mic\ cores, plus contour
maps for comparison with the near- and mid-IR data.  A more thorough
analysis of the 850\mic\ observations is beyond the scope of this
paper, and will be presented in a later work.


\begin{figure*}
  \epsfxsize=17.5cm
  \epsfbox{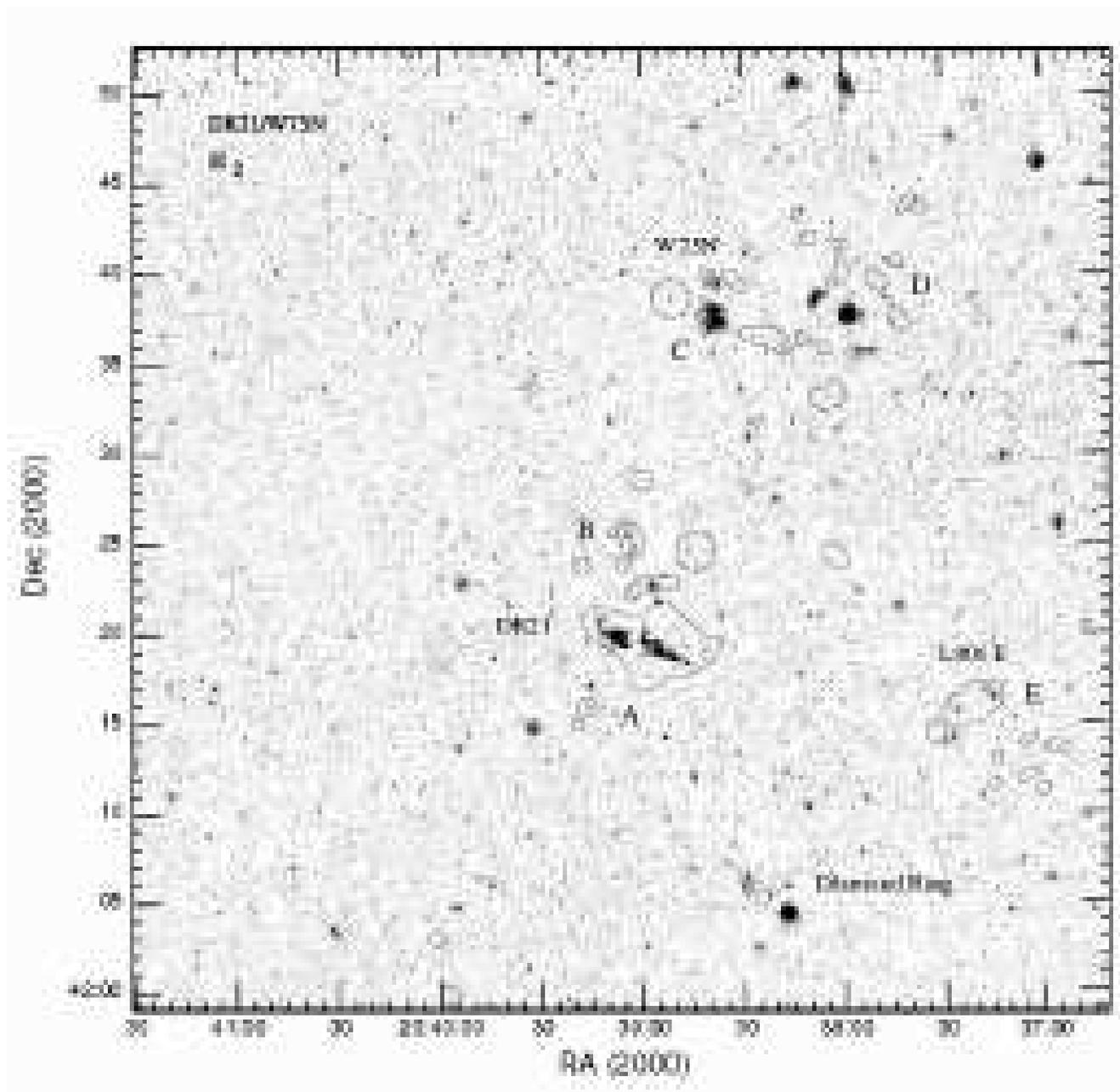}
\vspace*{0.3cm}
\caption[]
{A near-IR narrow-band \htwo\ image of the DR21 and W75N regions.  The figure
shows a full WFCAM tile, covering an area of roughly 0.8$\times$0.8
degrees. Areas were H$_2$ line emission is identified are outlined; 
the regions A, B, C, D and E are discussed separately in Sect.~4.}
\label{wfcam0}
\end{figure*}

\begin{figure*}
  \epsfxsize=17.5cm
  \epsfbox{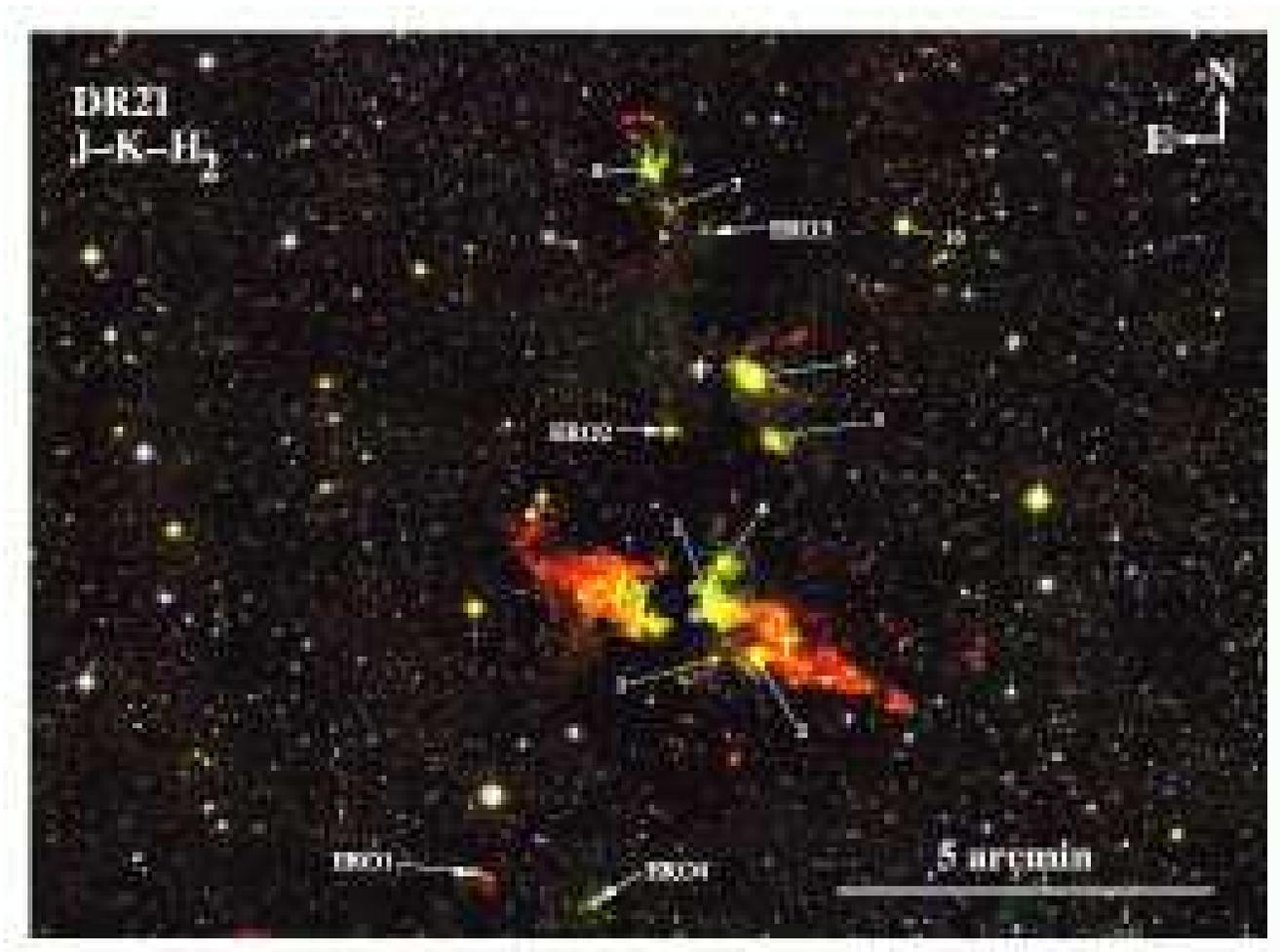}
\vspace*{0.3cm}
\caption[]
{A colour image of the DR21 region composed of observations in J
(blue), K (green) and H$_2$ (red). With this combination H$_2$ line
emission regions appear red, while embedded or background stars appear
yellow. The star and triangle mark the DR21 A-B-C and DR21 D cometary
H{\sc ii} regions; the cross indicates the location of the DR21(OH)
cluster of water, methanol and OH masers
\citep{lie97}.  EROs identified by \citet{mar04} are labeled; the infrared
``DR21-IRS'' sources discussed in the text are numbered.}
\label{wfcam1}
\end{figure*}

\begin{figure*}
  \epsfxsize=17.5cm
  \epsfbox{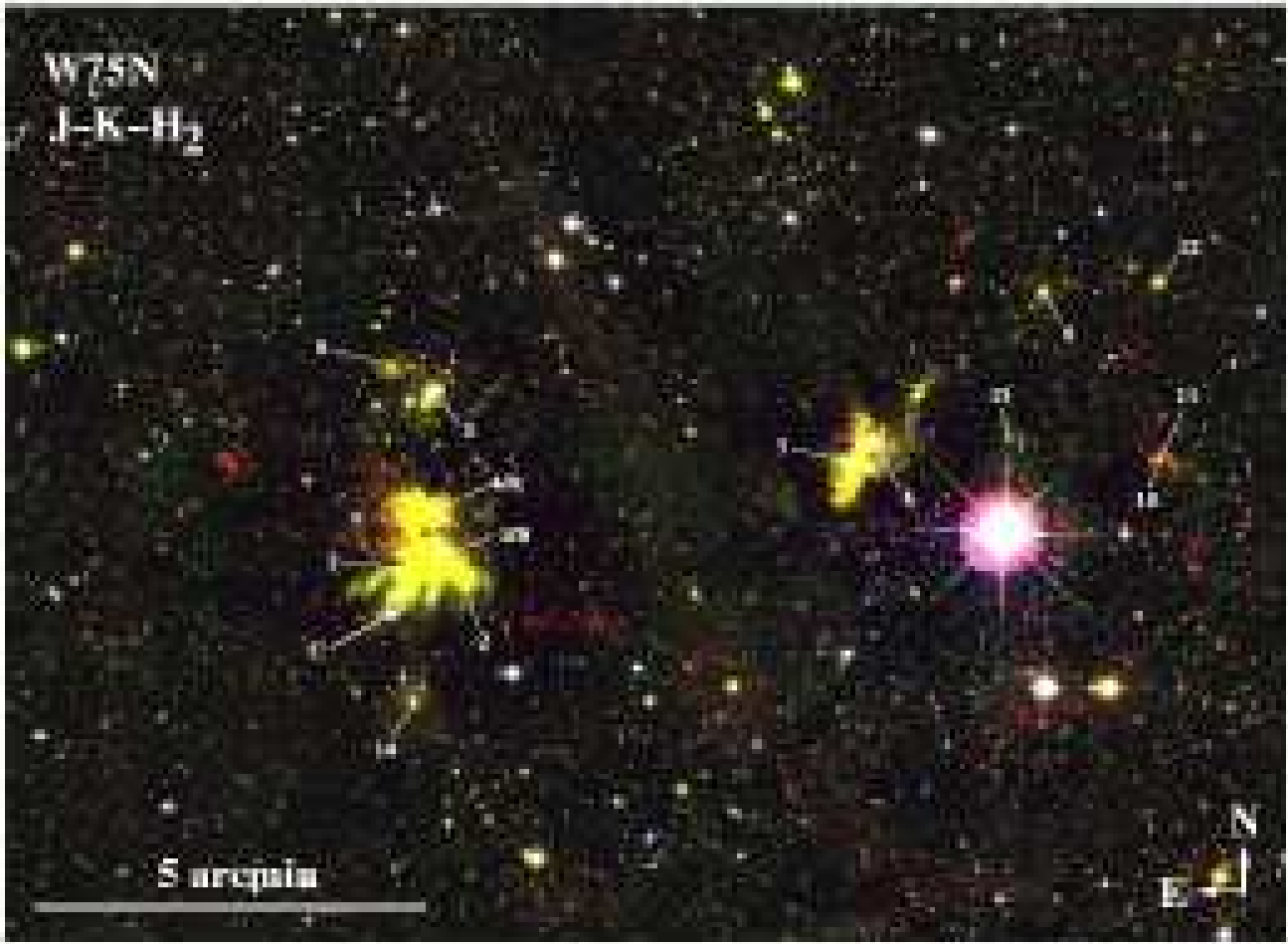}
\vspace*{0.3cm}
\caption[]
{Same as Fig.~\ref{wfcam1}, except for the region around W75N. The bright 
 infrared sources referred to as W75N-IRS~1, etc. in the text are
 numbered.  The letters A, B and C refer to the main H{\sc ii} regions.}
\label{wfcam2}
\end{figure*}

\begin{figure*}
\epsfxsize=17.5cm
  \epsfbox{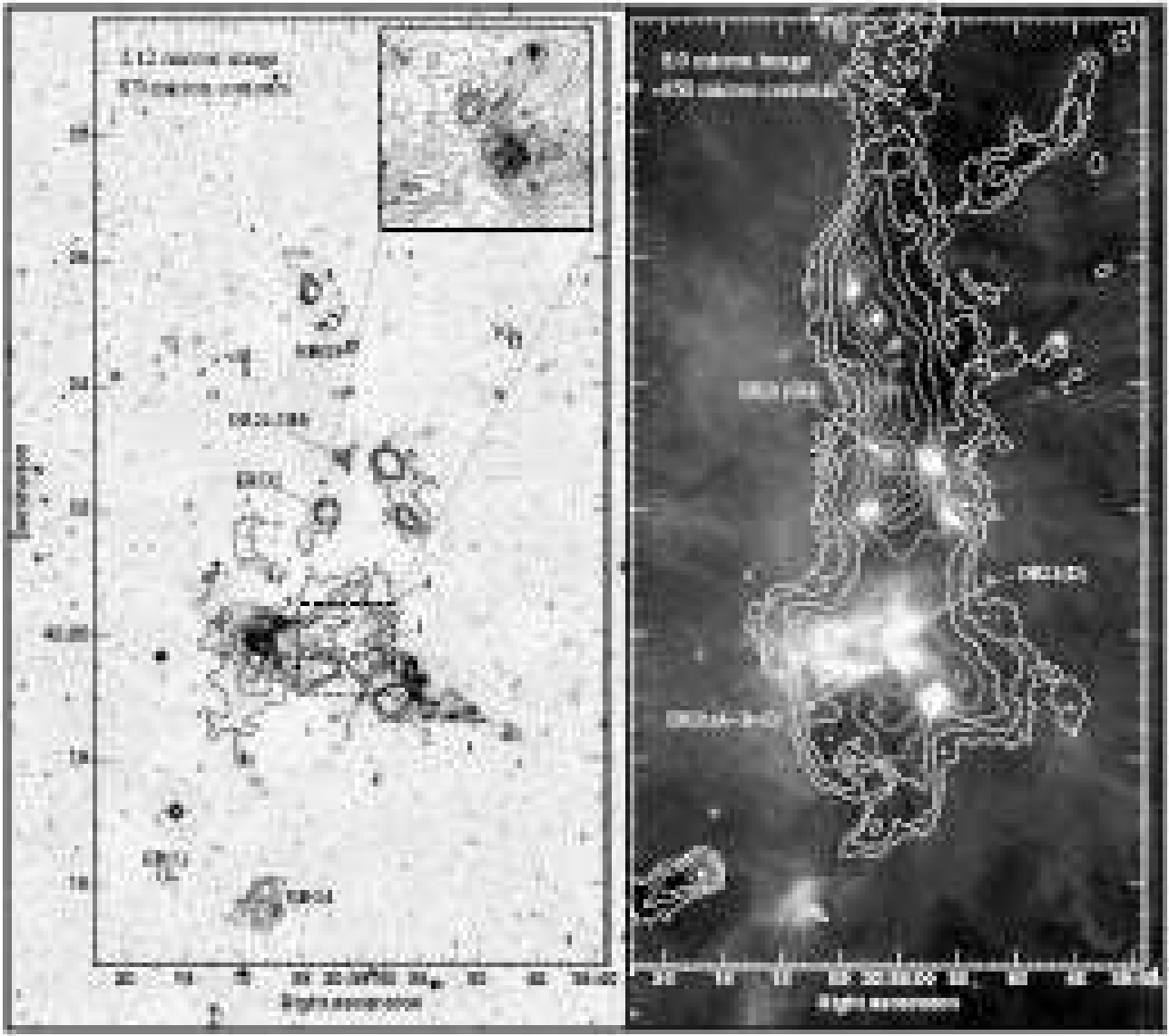}
\vspace*{0.3cm}
\caption{Left: WFCAM H$_2$ 2.12\mic\ image  of DR21 with 
Spitzer IRAC 8.0\mic\ contours overlayed. The contours pick out
only the brightest 8.0\mic\ peaks; lower contours associated with
diffuse PAH emission are not plotted.  Contour levels are  
 3.4, 5.1, 6.8, 10.2 mJy arcsec$^{-2}$ (black) and 
17.0, 30.6, 85.0, 165.0 mJy arcsec$^{-2}$ (white); inset, the same
contours are plotted but with thin and thick lines.  DR21-IRS sources are numbered.
Right: Spitzer IRAC 8.0\mic\ image with SCUBA 850\mic\ contours overlayed. 
Contour levels are 0.125, 0.25, 0.5, 1, 2, 4, 8 and 16 Jy/beam; the beam size
measures 14\arcsec.}
\label{spit2}
\end{figure*}

\begin{figure*}
\epsfxsize=17.0cm
  \epsfbox{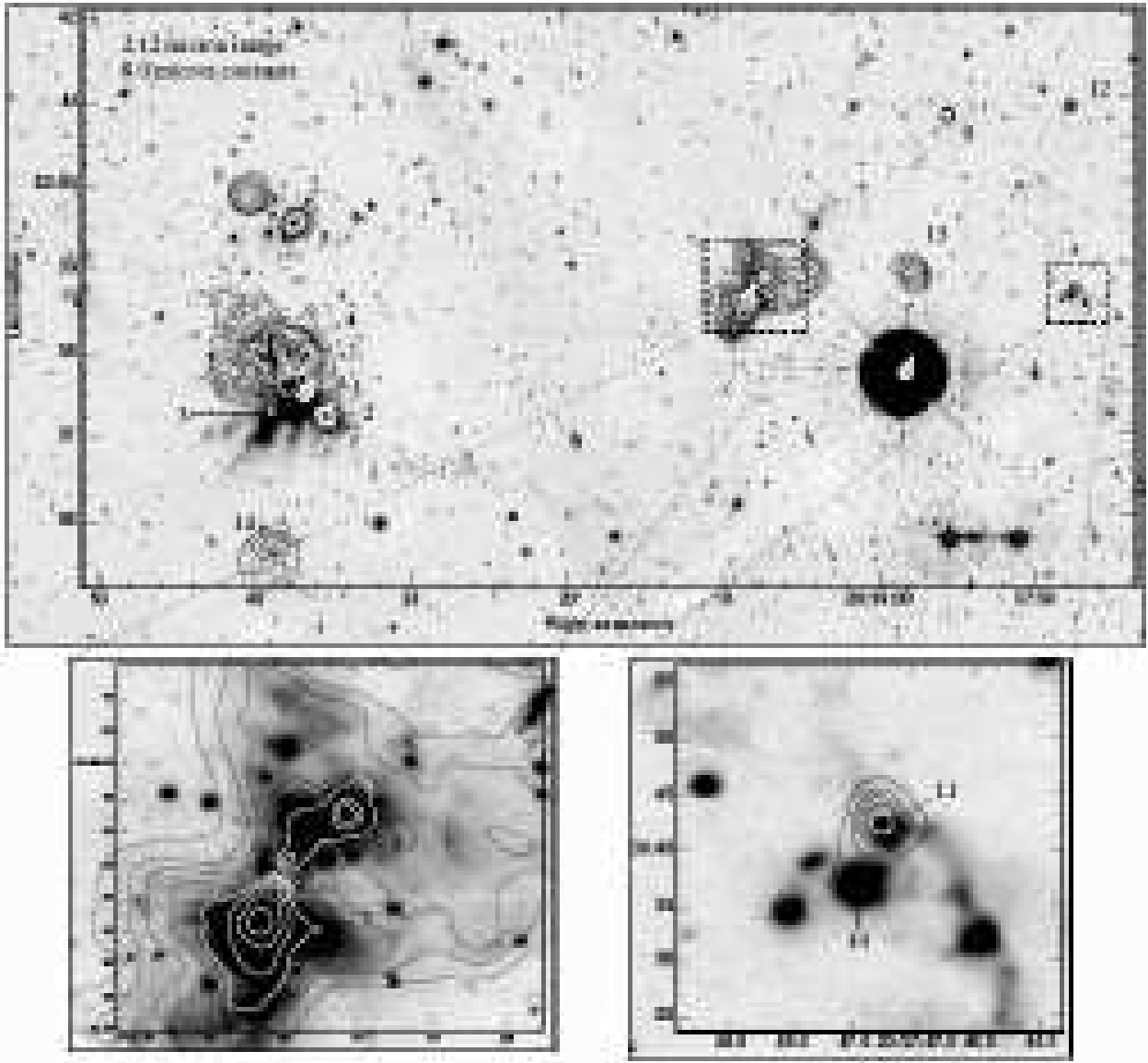}
\vspace*{0.3cm}
\caption{WFCAM H$_2$ 2.12\mic\ image  of W75N and the northwest, with 
Spitzer IRAC 8.0\mic\ contours overlayed.   Contour levels are  
 3.4, 5.1, 6.8, 10.2 mJy arcsec$^{-2}$ (black) and 
17.0, 30.6, 85.0, 165.0 mJy arcsec$^{-2}$ (white).
The W75N-IRS sources are numbered.
}
\label{spit3}
\end{figure*}

\begin{figure*}
\epsfxsize=15.0cm
 \epsfbox{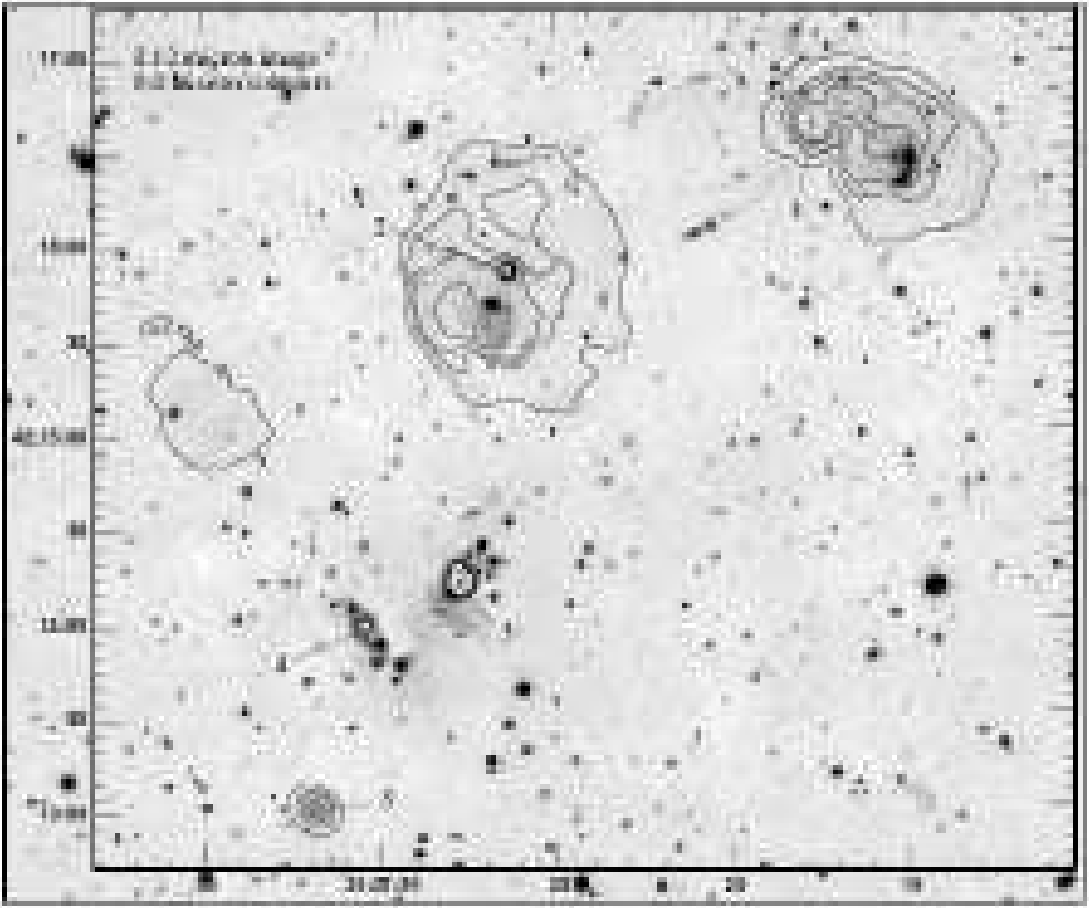}
\vspace*{0.3cm}
\caption{WFCAM H$_2$ 2.12\mic\ image of the eastern edge of L906 (west of
 DR21). Spitzer IRAC 8.0\mic\ contours are
 overlayed.
 Contour levels are 3.4, 5.1, 6.8, 10.2 mJy arcsec$^{-2}$ (black) and 
17.0, 30.6, 85.0, 165.0 mJy arcsec$^{-2}$ (white).
The L906E-IRS sources are numbered.
}
\label{spit4}
\end{figure*}

\begin{figure*}
\hspace*{0.1cm}
\epsfxsize=16.3cm
 \epsfbox{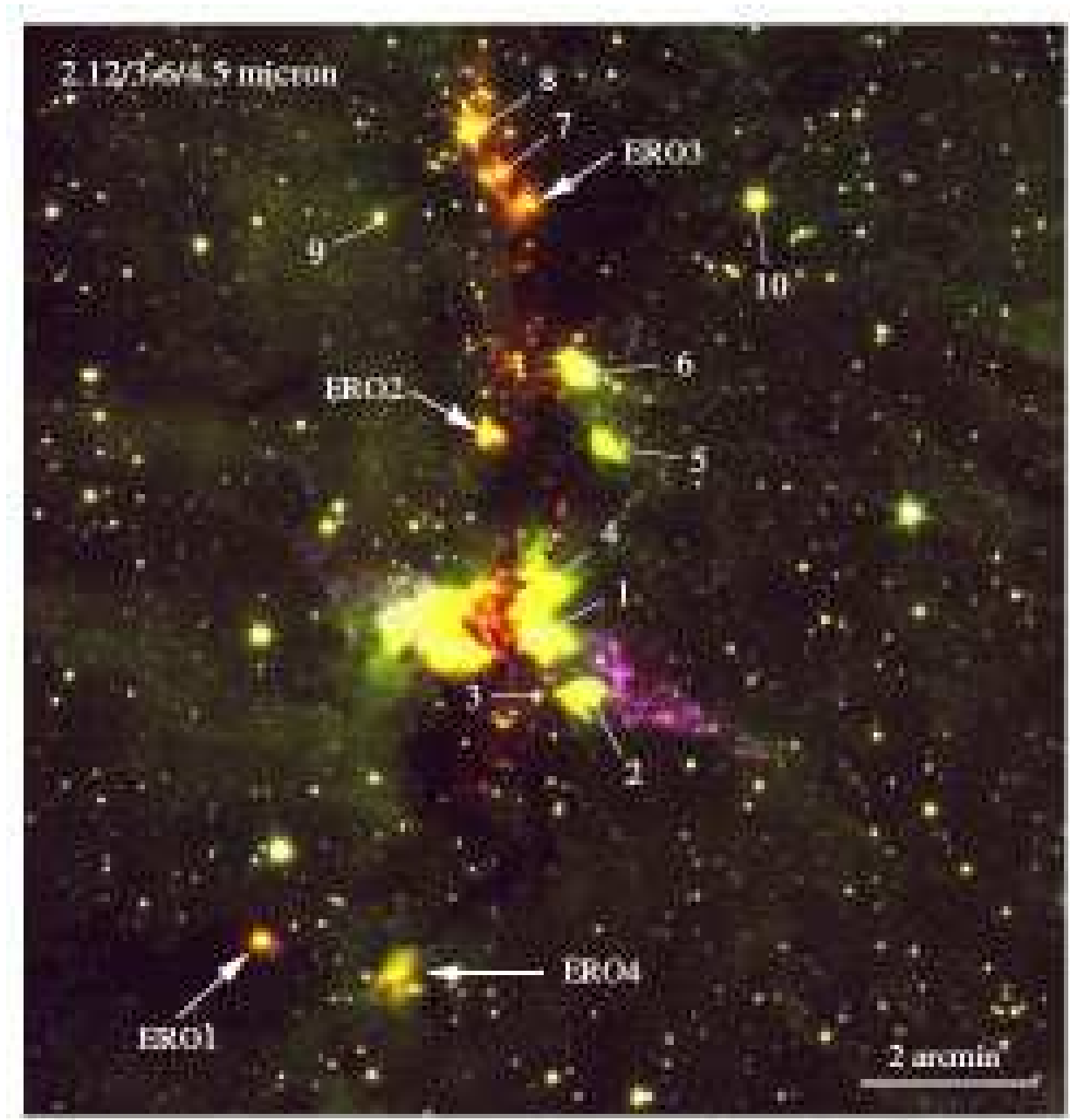}
\vspace*{0.3cm}
\caption[]
{Colour image constructed from WFCAM H$_2$ data (2.12\mic\ - blue)
and Spitzer IRAC images (3.6\mic\ - green; 4.5\mic\ - red).
EROs are labeled and DR21-IRS sources are numbered. }
\label{spit1}
\end{figure*}

\begin{figure*}
     \epsfxsize=17.5cm
 \epsfbox{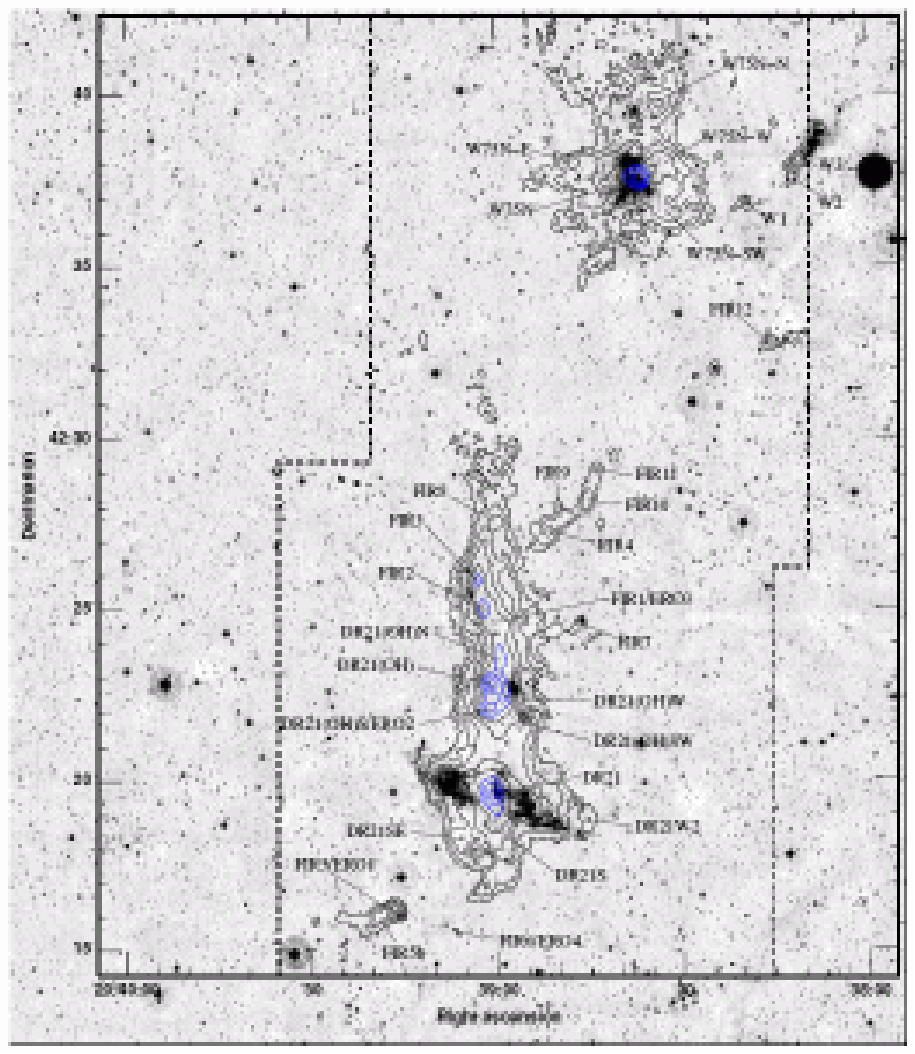}
\vspace*{0.3cm}
\caption[]
{WFCAM H$_2$ image of a portion of the DR21/W75 tile with,
superimposed, contours of the 850\mic\ emission. The SCUBA contours
represent 0.125, 0.25, 0.5, 1, 2, 4, 8 and 16 Jy/beam (the beam size
measures 14\arcsec\ at 850\mic ). The extent of the SCUBA map is
marked with dashed lines. }
\label{scu}
\end{figure*}


\section{Results}

\subsection{WFCAM imaging of DR21/W75}

Our H$_2$ 1-0S(1) (+ continuum) tile of the DR21/W75 region is
presented in Fig.1 (which shows the full field observed).  Regions
where H$_2$ line-emission features were identified have been outlined.
Although somewhat clustered around DR21 and W75N, these generally
follow the ridge of obscuration that links DR21 in the centre with
W75N in the north-northwest.  There is also a cluster of features
to the west-south-west of DR21, which may be unrelated to the main
high-mass star forming ridge.

Colour images of the regions around DR21 and W75N are shown in
Figs.~\ref{wfcam1} and \ref{wfcam2}.  J, K and H$_2$ mosaics were
scaled by exposure time, filter bandpass and filter transmission
before being registered and combined.  In the resulting figures
embedded sources can be easily distinguished from pure line-emission
features, i.e. the H$_2$ knots and bow shocks in jets and outflows.
The former appear yellow while the latter are bright red in colour.
Some of the yellow features in the DR21 lobes are due to scattered
continuum, while some may result from ``boosted'' K-band emission due
perhaps to enhanced vibrational excitation by fluorescence
\citep{fer97}.  Many of the features marked in Fig.~1 are seen in more
detail here; indeed, these colour images were used to search for H$_2$
features across this extensive region.  In Fig.~\ref{wfcam1} the
molecular ridge that runs through the middle of DR21 is evident as a
dark lane, largely devoid of point sources.  We mark on
Fig.~\ref{wfcam1} the ``Extremely Red Objects'', or EROs, identified
by \citet{mar04}, as well as the location of the DR21 A-B-C and D 
H{\sc ii} regions \citep{cyg03}, and the DR21(OH) cluster of water, OH and
methanol masers \citep{lie97}.  In Fig.~\ref{wfcam2} the H{\sc ii}
regions A, B and C in W75N \citep{has81} are labeled.  In both figures
K-band sources with bright 8.0\mic\ counterparts are numbered; these
are probably the most massive, embedded sources in the region. 
  
DR21 is famous for its H{\sc ii} regions.  It comprises four main radio
emission peaks, originally labeled A, B, C and D
\citep{har73,roe89}.  None of the radio peaks are detected in the 
near-IR \citep{dav96}, although the northern, compact source D is
observed at mid-IR wavelengths.
\citet{hsm05} point to this source as being the dominant source of
infrared luminosity in DR21.  They find that its total luminosity
exceeds that of an O8 ZAMS star and attribute the excess to accretion.
They estimate an extinction of $A_{\rm v} \sim 175$~mag toward 
source D, and suggest that because it is still accreting, it
could (at least in part) be responsible for the large-scale DR21
outflow.  However, at radio wavelengths H{\sc ii} region D is cometary
in appearance, opening outward toward the north \citep{cyg03}.  One
could argue that a north-south flow is more likely.  Either way, there
are no H$_2$ knots or jet-like features close to this source which one
might {\em confidently} associate with the H{\sc ii} region.

Situated $\sim$20\arcsec\ south of source D, radio sources A, B and C
appear to be peaks in a larger ``bow shock'' or ``champagne flow''
style cometary nebula \citep{cyg03}.  This nebula lies
$\sim$15\arcsec\ to the east of the cluster of near-IR sources in
DR21, within the dark lane that separates the two lobes of the DR21
outflow.  It is coincident with a dense molecular core (see Section
3.3), although it has no mid-IR counterpart.  Overall, the cometary
A-B-C nebula opens in the direction of the eastern lobe of the main
DR21 outflow. This might suggest a causal relationship between the
flow and the H{\sc ii} region.  However, the more collimated, western
lobe of the DR21 outflow is difficult to reconcile with the sharp,
western edge of the cometary H{\sc ii} region.  The powerful flow
would surely disrupt this H{\sc ii} region/molecular cloud boundary,
if the outflow central engine was situated {\em within} the cometary
nebula.

The complex of masers and mm-wave continuum peaks known as DR21(OH)
are found about 3\arcmin\ to the north of DR21 (A-B-C-D).  
DR21(OH) coincides with a chain of four water masers
\citep{for78,man92} spread northeast-southwest over about 20\arcsec.
Orthogonal to the water masers, about a dozen methanol masers
delineate the two lobes of the DR21(OH) outflow
\citep{pla90,kog98}. The water masers likely trace embedded young
stars and ultra-compact H{\sc ii} regions, including in this case the
outflow central engine, while the class I methanol masers probably
trace the interaction between the outflow and the ambient medium.
H$_2$ line emission is also excited in such outflow-ambient impact
zones, although we fail to detect clear evidence for H$_2$ emission in
our WFCAM data.  The absence of H$_2$ emission in the DR21(OH) outflow is
likely due to extinction.

W75N, shown in Fig.~\ref{wfcam2}, comprises three regions of radio
emission, W75N(A), W75N(B) and W75N(C) \citep{has81}.  A and B are
coincident with patches of bright near-IR nebulosity; the more evolved
source, W75N(A), is associated with a 2\mic\ point source. The overall
cluster harbours at least 30 IR sources with infrared excesses,
probably due to circumstellar disks \citep{moo91b,per06}. 

Collectively, W75N is associated with a $\sim10^3$\Msolar\ dense
molecular core \citep{moo91a}, which \citet{she03} resolve into nine
discrete millimetre continuum peaks spread over an area of about an
arcminute.  The brightest and most massive, MM~1, peaks a few
arcseconds to the northwest of IRS\,1 and is associated with H{\sc ii}
region W75N(B); W75N(A) is similarly associated with a continuum peak,
MM~5.  MM~1 contains at least four H{\sc ii} regions (a fifth is
associated with MM~5). Three of these were detected at 1.3~mm with the
VLA by \citet{tor97}. Of these, VLA~1 drives a radio jet in the
direction of the large-scale molecular outflow, although all three VLA
sources may drive independent flows \citep{she01,she03}.  All three
sources are associated with water maser emission
\citep{tor97}; VLA~1 is also surrounded by a compact cluster of OH
masers \citep{baa86,hut02}.  Each source is presumably in the
protostellar phase, their radio fluxes being consistent with early
B-type ZAMS stars.  Indeed, \citet{she04} assign spectral types of
B0-B2 to all five H{\sc ii} regions, W75N(A) being perhaps the most
evolved at B0.5.  Our H$_2$ observations of W75N reveal not only the
well-known H$_2$ features associated with W75N itself, but also
collimated jets to the north and west.

Lastly, roughly 15\arcmin -20\arcmin\ to the west of DR21 there is a
further region of active star formation that has to date attracted
little attention. The region is evident in the Spitzer data of
\citet{mar04}, although they do not discuss these features in any detail.  
 A search of the SIMBAD\footnote{The SIMBAD database 
is operated at CDS, Strasbourg, France.} database suggests that the
region is situated on the eastern edge of the small L906 dark cloud.
We find an abundance of H$_2$ line emission knots and jets, plus
regions of bright, nebulous, mid-IR emission and a number of highly
reddened sources (Paper II).  We use the suffix L906E for the sources
in this region.

Notably, no H$_2$ flows were found in the Diamond Ring region
(Fig.~\ref{wfcam0}), which is consistent with this area being more
evolved than DR21, W75N and L906E.


\subsection{Comparison with Spitzer data}

\subsubsection{The young stars}

In Figs.~\ref{spit2}--\ref{spit4} we compare our WFCAM observations of
DR21/W75 with the archival Spitzer 8.0\mic\ observations first
discussed by \citet{mar04} and \citet{per06}.  These longer-wavelength
data are potentially very useful for tracing luminous, embedded
sources.  Photometry of the brighter 8.0\mic\ point sources is given
in Table~\ref{irs}.

The EROs identified by \citet{mar04} in DR21 are bright, reddened 8\mic\
sources with spectral indices ($\alpha = d {\rm log}\lambda
F_{\lambda} / d {\rm log}\lambda$) between 2.2\mic\ and 8.0\mic\ in
the range 1.8 to 3.6, much larger than the $\alpha > 0.3$ lower limit
for Class I protostars \citep{lad87}.  ERO~1 lies at the western edge
of an ``infrared dark cloud'' that extends eastward for at least
2\arcmin\ \citep{mar04}; the cloud is evident as a region devoid of
stars in Fig.~\ref{wfcam1}.  In the 850\mic\ data
described below the cloud is seen as two peaks (FIR~5 and FIR~5b)
enveloped in an easterly elongation of the dust emission.

The other extremely red objects, ERO~2, ERO~3 and ERO~4, all lie along
the north-south chain of dense cores that runs through the centre of
DR21.  In Table~\ref{irs} we list the Spitzer magnitudes of these and
many other bright 8.0\mic\ sources in DR21/W75: in Fig.~\ref{spit2}
(around DR21), Fig.~\ref{spit3} (near to W75N) and Fig.~\ref{spit4}
(around L906E), these sources are labeled DR21-IRS~1, DR21-IRS~2,
etc., W75N-IRS~1, W75N-IRS~2, etc, and L906E-IRS~1, L906E-IRS~2, etc,
respectively (the same sources are labeled in the near-IR images in
Figs.~\ref{wfcam1} and \ref{wfcam2}).  These ``IRS'' sources are
generally bright 8.0\mic\ point sources or luminous, reddened K-band
sources associated with bright 8.0\mic\ emission (surface brightness
$>$7~mJy arcsec$^{-2}$).  All appear as peaks or discrete sources in
Figs.~\ref{spit2}-\ref{spit4}.  Spectral indices, $\alpha$, derived
from linear fits to the Spitzer photometry between 3.6\mic\ and
8.0\mic\ (as described in Paper II), are also given in
Table~\ref{irs}.  We do not correct for extinction since this is not
know a priori for each source, although it is worth noting that, at a
distance of 3~kpc the interstellar extinction to DR21/W75 may be as
large as $A_{\rm v}\sim8.8$~mag \citep{jos05}.
Extinction will have the effect of increasing the range of $\alpha$
values (Paper II).  For targets with photometry in only two IRAC bands
we do not derive values for $\alpha$, since the 8.0\mic\ magnitudes
may be compromised by deep silicate absorption in these luminous,
embedded sources
\citep[e.g][]{gib00}.

Most of the targets in Table~\ref{irs} are very red.  Indeed, when
considering all of the sources, the EROs identified by \citet{mar04}
do not seem to be particularly unique.  We find many other examples of
extremely red objects. DR21-IRS~5 and L906E-IRS~1 are particularly
noteworthy, having $\alpha$ (uncorrected for extinction) of $\sim 4.1$.
UCH{\sc ii} region DR21(D) is also very red, as is DR21(OH).  For the
EROs, the spectral indices in Table~\ref{irs} differ somewhat from
those reported by \citet{mar04}, probably because different photometry
apertures (and sky annuli) were used with these moderately extended
sources.  Our Spitzer photometry in the W75N regions matches that
reported by \citet{per06}, who use a similar photometry aperture to
ourselves, reasonably well.

The mid-infrared sources in the W75N region are shown in
Fig.~\ref{spit3}.  The region associated with W75N(A) and (B) is very
complex.  Ground-based and Spitzer mid-IR photometry of the
sources in the immediate vicinity of W75N have been discussed in
detail by \citet{per06}, although there are additional mid-IR sources
to the north and west close to bright infrared nebulae and regions of
H$_2$ line emission. \citet{per06} identify mid-IR counterparts to the
near-IR sources W75N-IRS\,1, 2 and 3, as well as the VLA\,3 UCH{\sc ii}
region in MM~1.  They derive an extinction of $A_{\rm v} \sim 90$ to
VLA\,3. 

Much like the IRS sources near DR21, around W75N and to the west, the
combined WFCAM and Spitzer data reveal nebulous sources which harbour
a luminous O or early-B star within a compact cluster of less-luminous
sources (e.g. W75N-IRS~1 to W75N-IRS~5, W75N-IRS~7 and probably
W75N-IRS~8), as well as examples of single intermediate-mass sources
within an almost spherical shell of reflected near-IR light and
8.0\mic\ PAH emission (e.g.  W75N-IRS~6, IRS~13 and IRS~14).
 
In Fig.~\ref{spit4} we show WFCAM 2.12\mic\ and Spitzer 8.0\mic\ data
of star forming regions to the west of DR21. The Spitzer data reveal a
chain of three spherical cores. As one moves westward along this chain
the cores become less diffuse and more centrally peaked at 8.0\mic :
note that the eastern core in Fig.~\ref{spit4} has no 8.0\mic\ point
source, while L906E-IRS~2 is markedly fainter than L906E-IRS~1. IRS~1
is extremely red (particularly in comparison to L906E-IRS~2;
Table~\ref{irs}), and is associated with extensive H$_2$ filaments
that probably delineate at least two jets (discussed further in
Section A5). If we also include the cluster of YSOs associated with
L906E-IRS~3/IRS~4, then we probably have an evolutionary sequence,
with L906E-IRS~1 representing the youngest and L906E-IRS~3/IRS~4 the
most evolved region.
 
The bright Spitzer sources and the clustering properties of the
embedded stellar population are discussed in more detail in Paper II.

\subsubsection{The outflows}

In addition to the embedded young stars, Spitzer also detected the
main DR21 outflow at 4.5\mic . In high-mass star forming regions like
DR21/W75, distinguishing line-emission jet features from background
diffuse dust and PAH emission regions is difficult, even in the
relatively PAH-free 4.5\mic\ band.  With the broad IRAC bands one is
not able to ``continuum-subtract'' the data.  One must therefore use
colour images as a diagnostic tool.  We show such an image in
Fig.~\ref{spit1}.

As noted by \citet{mar04} and more recently by \citet{hsm06}, the main
DR21 outflow is detected by Spitzer.  Morphologically, the flow is
very similar in images at 2.12\mic\ and 4.5\mic . Indeed,
\citet{hsm06} overlay contours of the H$_2$ 1-0S(1) emission of
\citet{dav96} onto a 4.5\mic\ image and find an almost exact
correlation between the emission-line features in the two bands.  Note
also that the flow is consistently the same colour (pink) in
Fig.~\ref{spit1}.  Although the IRAC filter at 4.5\mic\ encompasses
molecular and atomic emission lines from CO (v=1-0 at 4.45-4.95\mic )
and HI Br$\alpha$ (at 4.052\mic ), the unchanging colour along the
DR21 outflow in Fig.~\ref{spit1} suggests that the bulk of the
4.5\mic\ emission is in fact from the H$_2$ 0-0S(9) line (at 4.694\mic
).  This conclusion is supported by ISO spectroscopy of portions of
the DR21 outflow \citep{smi98,hsm06}.

In shock models, H$_2$ emission in outflows is predicted to be
particularly strong in the 4.5\mic\ band, 5--14 times higher than in
the 3.6\mic\ band \citep{smi05}. \citet{hsm06} find that {\em in
outflows} $\sim$50\% of the flux in the IRAC 3.6\mic , 4.5\mic\ and
5.8\mic\ bands is due to H$_2$ line emission.  \citet{smi98} propose
excitation of the H$_2$ emission in multiple C-type bow
shocks. \citet{hsm06} also note a conspicuous lack of CO, Br$\alpha$
and PAH emission in the ISO spectra.  

Most of the ``red'' H$_2$ features evident in our WFCAM images
(Figs.~\ref{wfcam1} and \ref{wfcam2}) are detected by Spitzer, albeit
at lower spatial resolution.  In Fig.~\ref{spit1} one or two features
along the ridge appear somewhat ``redder'' than other jet features,
probably because of increased extinction (note in particular the knot
$\sim$1\arcmin\ north of DR21-IRS~4). This may also explain the brown
colouration at the end of the western DR21 flow lobe, although
\citet{hsm06} -- who also note this subtle change in colour -- mention
that a reduced pre-shock velocity or increased pre-shock density may
 have the same effect.

The Spitzer 5.8\mic\ band covers emission from the strong [FeII] line
at 5.34\mic .  However, upon examining the 5.8\mic\ images (not shown)
we found it impossible to distinguish line emission features
associated with the outflow from the more intense, filamentary
6.2\mic\ PAH emission that pervades the 5.8\mic\ data (the 3.6\mic\
and 8.0\mic\ bands likewise include bright PAH emission at 3.3\mic\
and 7.7\mic ). Suffice to say that no {\em bright} [FeII] component
was found in the main DR~21 outflow.


 \begin{table*}
 \begin{tiny}
 \centering
 \begin{minipage}{175mm}
  \caption{Catalogue of bright 8\mic\ sources in DR21/W75}
  \begin{tabular}{@{}lllcccccll@{}}
  \hline
Ref$^1$ &  RA$^2$     & Dec$^2$     & [3.6] & [4.5] & [5.8] & [8.0] & $\alpha$ & Name & Descriptive comment   \\
    &  (2000.0)   &(2000.0)     & \\
 \hline

1  & 20:39:16.72  & 42:16:09.1  & 8.12 & 7.06 & 5.49 & 4.82 & 1.1 & ERO1       & K-band/8\mic\ point source (``a'' in Table~\ref{sources}) \\
2  & 20:39:02.86  & 42:22:00.2  & 10.48& 9.86 & 7.29 & 5.58 & 3.2 & ERO2       & Diffuse 8\mic\ extending $\sim$5\arcsec\ W of K-band star  \\
3  & 20:39:00.47  & 42:24:36.6  & 9.74 & 8.00 & 6.87 & 5.93 & 1.5 & ERO3       & K-band/8\mic\ point source \\
4  & 20:39:08.33  & 42:15:44.9  & 11.25& 10.48& 9.32 & 7.80 & 1.2 & ERO4c      & K-band point source in NE 8.0\mic\ arc \\
5 & 20:39:07.28  & 42:15:34.8  & 12.06& 11.38& 8.65 & 6.90 & 3.5 &  ERO4s      & K-band point source in SW 8.0\mic\ arc \\ \\

6  & 20:39:00.05  & 42:19:36.6  &  --  &  --  & --   & --   & --  & DR21-IRS1  & No distinct 8\mic\ peak \\
7  & 20:38:56.91  & 42:18:58.8  &  --  &  --  & --   & --   & --  & DR21-IRS2  & Diffuse 8\mic\ peak $\sim$3\arcsec N of K-band star \\
8  & 20:38:59.83  & 42:18:58.1  & 10.74& 10.18& 9.44 & 8.53 & -0.3& DR21-IRS3  & Faint K-band/8\mic\ point source \\
9  & 20:38:59.32  & 42:20:16.1  & 9.82 & 9.14 & 8.02 & 6.33 & 1.2 & DR21-IRS4  & Faint K-band/8\mic\ source with extended emission \\
10  &20:38:55.85$^*$& 42:21:54.1$^*$&12.06&11.37&8.32 & 6.47 & 4.1& DR21-IRS5  & Diffuse 8\mic\ peak; no K-band point source counterpart \\
11 & 20:38:57.19  & 42:22:41.1  & 7.40 & 6.62 & 5.35 & 4.13 & 1.0 & DR21-IRS6  & Bright K/8\mic\ point source (``e'' in Table~\ref{sources}) \\
12 & 20:39:02.00  & 42:24:59.2  & 9.22 & 6.92 & 5.46 & 4.54 & 2.4 & DR21-IRS7  & Bright 8\mic\ source with faint K-band counterpart \\
13 & 20:39:03.72  & 42:25:29.7  & 7.96 & 6.62 & 5.17 & 4.46 & 1.3 & DR21-IRS8  & Bright K/8\mic\ source  \\
14 & 20:39:09.49  & 42:24:26.8  & 9.32 & 8.51 & 7.86 & 7.03 & -0.2& DR21-IRS9  & K/8\mic\ point source  \\
15 & 20:38:46.37  & 42:24:39.7  & 7.63 & 6.91 & 5.86 & 5.25 & -0.0& DR21-IRS10 & Slightly extended K/8\mic\ point source \\ \\

16 & 20:38:36.76  & 42:37:29.4  &  --  &  --  &  --  &  --  &  -- & W75N-IRS1 & SE source of three 8\mic\ peaks \\
17 & 20:38:35.37  & 42:37:13.4  &  --  &  --  &  --  &  --  &  -- & W75N-IRS2 & Nebulous source at K and 8\mic  \\
18 & 20:38:38.82  & 42:37:17.6  & 8.96 & 8.30 & 7.65 & 6.32 & 0.2 & W75N-IRS3 & K-band/8\mic\ point source \\
19 & 20:38:37.73  & 42:37:59.4  &  --  &  --  &  --  & --   & --  & W75N-IRS4 & Diffuse 8\mic\ emission enveloping K-band point source \\
20 & 20:38:37.29  & 42:39:33.0  & 10.08& 9.62 & 7.21 & 5.44 & 2.9 & W75N-IRS5 & Bright K/8\mic\ point source  \\
21 & 20:38:40.16  & 42:39:53.5  &  --  &  --  &  --  &  --  &  -- & W75N-IRS6 & K-band point source in shell of 8\mic\ emission  \\
22 & 20:38:08.27  & 42:38:36.3  & 7.09 & 6.26 & 4.80 & 4.26 & 0.6 & W75N-IRS7 & K-band/8\mic\ point source (``m'' in Table~\ref{sources}) \\ 
23 & 20:38:07.10  & 42:38:52.1  &  --  & 7.38 &  --  & 5.39 &  -- & W75N-IRS8$^3$ & K-band/8\mic\ point source  \\
24 & 20:37:55.40  & 42:40:46.9  &  --  & 8.27 &  --  & 5.76 &  -- & W75N-IRS9$^3$ & K-band/8\mic\  point source\\
25 & 20:37:47.48  & 42:38:37.0  &  --  & 9.71 &  --  & 8.00 &  -- & W75N-IRS10$^3$& K-band/8\mic\  point source \\
26 & 20:37:47.25  & 42:38:42.0  &  --  & 10.04&  --  & 6.95 &  -- & W75N-IRS11$^3$& K-band/8\mic\  point source \\
27 & 20:37:47.35  & 42:40:53.0  &  --  & 7.69 &  --  & 7.19 &  -- & W75N-IRS12$^3$& K-band/8\mic\  point source \\
28 & 20:37:57.77  & 42:38:54.0  &  --  & 11.06&  --  & 6.78 &  -- & W75N-IRS13$^3$& K-band/8\mic\  point source \\
29 & 20:38:38.69  & 42:35:34.2  & 11.69& 11.31& 10.21& 8.79 & 0.6 & W75N-IRS14& K-band point source in shell of 8\mic\ emission \\ \\

30 & 20:37:17.67  & 42:16:37.4  & 10.62& 10.03& 6.83 & 5.04 & 4.1 & L906E-IRS1& K/8\mic\ point source (``n'' in Table~\ref{sources})\\
31 & 20:37:26.35  & 42:15:51.9  & 8.58 & 8.00 & 7.44 & 7.02 & -1.0& L906E-IRS2& K/8\mic\ point source in shell of 8\mic\ emission \\
32 & 20:37:27.73  & 42:14:13.1  & 8.72 & 7.78 & 6.53 & 4.93 & 1.6 & L906E-IRS3& Bright K/8\mic\ point source \\
33 & 20:37:30.50$^*$& 42:13:59.2$^*$&11.01&9.09& 7.67& 6.67 & 2.1 & L906E-IRS4& Bright 8\mic\ point source; no K counterpart \\
34 & 20:37:31.91  & 42:13:01.5  & 11.64& 11.03& 8.39 & 6.59 & 3.4 & L906E-IRS5& Bright K/8\mic\ point source \\
35 & 20:36:57.90  & 42:11:29.7  &  --  & 10.46&  --  & 8.92 &  -- & L906E-IRS6$^3$& Nebulous K-band source(s) \\ \\

36 & 20:39:01.27$^*$& 42:19:54.1$^*$&10.21&7.52&5.53 & 3.98 & 4.2 & DR21-D   & Very bright 8\mic\ point source; no K counterpart \\
37 & 20:39:01.01$^*$& 42:22:50.2$^*$&11.80&10.74&8.13& 6.27 & 3.8 & DR21(OH) & 8\mic\ point source; no K counterpart \\

 \hline
 \label{irs}
 \end{tabular}
 \smallskip \\
 $^1$Reference number used in the colour-colour diagrams in Paper II.  These are {\em not} 
     the ``IRS'' designations used here in the figures.\\
 $^2$Coordinates of K-band point source counterparts to bright 8\mic\ features in DR21/W75.  
     Coordinates marked with a $^*$ have no K-band source, so the 8\mic\ peak position is 
     listed. \\
 $^3$These targets were not observed in all IRAC bands.
 
 \end{minipage}
 \end{tiny}
 \end{table*}


 \begin{table*}
 \centering
 \begin{minipage}{175mm}
  \caption{Characteristics of the brightest 850\mic\ cores in DR21/W75}
  \begin{tabular}{@{}lcccccl@{}}
  \hline
Source        &   RA     &   Dec         & Flux$^{a}$ & Size$^{b}$ & P.A.$^{b}$ & 8\mic\ counterpart \\
              & (2000.0) & (2000.0)      &  (Jy)      &  (arcsec)  & (degs)     &  \\
  \hline
DR21$^{c}$      & 20 39 00.91 & 42 19 35.9 & 54.5 & 27 $\times$16     & 13  & No mid-IR counterpart \\
DR21S           & 20 39 00.70 & 42 18 11.5 & 0.65 & 6.4$\times$3.0    & -85 & No mid-IR counterpart \\
DR21SE          & 20 39 05.99 & 42 18 19.4 & 3.1  & 29$\times$25      & -59 & No mid-IR counterpart \\
DR21W1    & 20 38 54.50 & 42 19 14.0 & 6.5$^{d}$  & 38$\times$18.4    & 18  & $\sim$30\arcsec\ WNW of DR21-IRS~2; western DR21 flow \\
DR21W2          & 20 38 45.36 & 42 18 50.0 & 0.23 & 15.5$\times$9.7   & -28 & No mid-IR counterpart \\
DR21E     & 20 39 08.25 & 42 19 55.9 & 0.36$^{d}$ & 5.8$\times$3.3    & -20 & Diffuse 8\mic\ emission; eastern DR21 outflow lobe \\ \\

DR21(OH)$^{e}$  & 20 39 00.80 & 42 22 49.0 & 32.8 & 14.1$\times$12.4  &  57 & $\sim$3.5\arcsec\ WSW of 8\mic\ peak \\
DR21(OH)SW      & 20 38 57.20 & 42 21 47.4 & 2.1  & 36$\times$11.4    & 60  & $\sim$15\arcsec\ SE of DR21-IRS~5 \\
DR21(OH)S       & 20 39 01.36 & 42 22 06.7 & 16.6 & 21.2 $\times$12.5 & -68 & ERO~2; $\sim$15\arcsec\ WNW of 8\mic\ peak \\
DR21(OH)W       & 20 38 59.00 & 42 22 23.9 &  9.0 & 16.2 $\times$12.0 & 17  & No mid-IR counterpart \\
DR21(OH)N$^{f}$ & 20 38 59.52 & 42 23 44.6 &  8.6 & 42$\times$12      & 16  & No obvious mid-IR counterpart \\ \\

FIR~1           & 20 39 00.20 & 42 24 37.0 &  4.4 & 29.8$\times$8.3   & -85 & ERO~3; $\sim$2.0\arcsec\ W of elongated 8\mic\ peak \\
FIR~2           & 20 39 02.16 & 42 25 01.3 &  9.2 & 22.5$\times$11.9  & 12  & $\sim$2.5\arcsec\ NE of elongated 8\mic\ pea \\
FIR~3           & 20 39 03.11 & 42 25 51.6 &  4.4 & 12.2$\times$5.7   & -13 & Coincides with four faint 8.0\mic\ sources \\
FIR~4           & 20 38 51.52 & 42 27 15.6 & 1.3 & 27$\times$15.9     & 21  & No mid-IR counterpart  \\
FIR~5           & 20 39 16.81 & 42 16 11.3 &  3.2 & 15.1$\times$12.1  & -75 & ERO~1; coincides with pair of 8.0\mic\ sources \\
FIR~5b          & 20 39 19.28 & 42 16 02.0 &  1.0 & 13.2$\times$8.6   & -57 & No mid-IR counterpart  \\
FIR~6           & 20 39 06.98 & 42 15 34.6 &  1.0 & 36$\times$19      &  15 & ERO~4; $\sim$4 WSW of K-band/8.0\mic\ source ERO~4s\\
FIR~7           & 20 38 46.54 & 42 24 37.1 & 0.40 & 7.8$\times$6.3    & -25 & $\sim$3\arcsec\ SE of  DR21-IRS~10 \\
FIR~8           & 20 39 00.09 & 42 27 32.0 & 1.7  & 22$\times$20      &  -- & No mid-IR counterpart  \\
FIR~9           & 20 38 49.10 & 42 27 40.3 & 0.63 & 15.0$\times$12.8  & -36 & No mid-IR counterpart  \\
FIR~10          & 20 38 45.50 & 42 28 10.3 & 2.2  & 52$\times$14.8    & -22 & No mid-IR counterpart  \\
FIR~11          & 20 38 43.76 & 42 29 03.5 & 0.62 & 19.3$\times$15.5  & -48 & No mid-IR counterpart  \\
FIR~12          & 20 38 15.40 & 42 32 44.8 & 0.62 & 16.7$\times$15.2  & -59 & No mid-IR counterpart  \\ \\


W75N$^{g}$      & 20 38 36.41 & 42 37 34.2 & 26.4 & 11.8 $\times$8.8  & 10  & Coincides with elongated 8\mic\ peak, to within $\sim$1\arcsec .\\
W75N-E          &  20 38 47.28 & 42 38 02.0 & 1.1 & 22$\times$6.3     & -83 & No mid-IR counterpart \\
W75N-N          &  20 38 33.18 & 42 39 45.7 & 3.8 & 31$\times$27      & -22 & No mid-IR counterpart, though near outflow sources \\
W75N-W          &  20:38:31.02 & 42:37:47.6 & 2.3 & 39$\times$10      &  81 & No mid-IR counterpart \\
W75N-SW         &  20 38 31.02 & 42 36 27.1 & 4.0 & 43$\times$33      & -85 & No mid-IR counterpart \\
W75N-W1         &  20:38:19.47 & 42:36:52.5 & 0.2 & 11$\times$7.6     & -71 & $\sim$5\arcsec\ S of *\mic\ peak \\
W75N-W2         &  20:38:11.82 & 42:37:34.6 & 0.6 & 21$\times$3       & -20 & No mid-IR counterpart \\
W75N-W3   & 20:38:10.65  & 42:38 04.6 &$\sim$1.0$^{h}$ & 9.1$\times$8.7 & --& $\sim$40\arcsec\ SE of W75N-IRS~7 \\

  \hline
 \label{scuba}
 \end{tabular}
 \smallskip \\
 $^a$Integrated flux at 850$\mu$m. \\
 $^b$Deconvolved size of the core and position angle of the long axis 
    (measured east of north).\\
 $^c$This primary peak is midway between H{\sc ii} regions DR21 A,B and C \citep{cyg03}. \\
 $^d$Probably includes CO J=3-2 line emission from the outflow; the SCUBA filter is centred
     at 353~GHz (passband $\sim$30~GHz), while the rest frequency of CO 3-2 is 346~GHz. \\
 $^e$Resolved by \citet{man91} into two cores at 2.7~mm. \\
 $^f$Resolved by \citet{cha93a} into two cores at 1.3~mm, N1 \& N2. \\
 $^g$Sub-mm core coincides with H{\sc ii} region W75N(B), which is further resolved into 
  three UC H{\sc ii} regions VLA1, VLA2, \& VLA 3 \citep{tor97}. \\
 $^h$Flux uncertain because this source is near the edge of the SCUBA map. \\
 \end{minipage}
  \end{table*}


\subsection{Comparison with SCUBA data}
 
In Fig.~\ref{scu} we overlay contours of 850\mic\ emission on to our
H$_2$ image.  The 850\mic\ scan map covers a region approximately
12\arcmin$\times$29\arcmin\ in size, encompassing both DR21, W75N and
the region in between. A comparison of SCUBA and Spitzer 8.0\mic\ data
is also made in Fig.~\ref{spit2}. Parameters derived for the major cores
in our map from Gaussian fitting of our flux-calibrated 850\mic\ data
are given in Table~\ref{scuba}.  The integrated flux is the
sky-subtracted value within the elliptical aperture defined in the
table.  The relationship (if any) to 8.0\mic\ point sources is also
noted, although we find measurable offsets between the SCUBA peaks and
mid-IR sources in almost all cases.  This is probably due to the fact
that the SCUBA cores, and in some cases 8.0\mic\ peaks, encompass
multiple sources.  Clearly, care needs to be taken when comparing
source fluxes at mid-IR and far-IR wavelengths.

The DR21 region comprises three groups of
molecular cores, the DR21 H{\sc ii} region itself, DR21(OH) and, a
further 3\arcmin\ to the north, a chain of cores labeled FIR~1,
2 and 3 by \citet{cha93a}.  New molecular cores are labeled FIR~4 to
FIR~12, while sub-components within the DR21 and DR21(OH) regions are
identified based on their relative locations.  The brightest 850\mic\
peak, labeled DR21 in Fig.~\ref{scu} (and Fig.~\ref{spit2}), has no
mid-IR counterpart, though it coincides with the H{\sc ii} regions
A-B-C \citep{cyg03}.  The 850\mic\ peak towards DR21(OH) coincides
precisely with the  3.1\,mm continuum source MM1 observed by
\citet{lie97}.

As already noted, ERO~1 is detected at 850\mic . However, neither
ERO~2 nor ERO~ 3 coincide precisely with an
850\mic\ peak; ERO~2 in particular is offset $\sim$15\arcsec\ from the
nearest SCUBA core, DR21(OH)S.  At 8.0\mic\ ERO~4
comprises three 8\mic\ sources.  Close examination of
Fig.~\ref{spit2} suggests that at least two of these emission features
-- ERO~4c and ERO~4s in the nomenclature of \citet{mar04} -- are
actually the illuminated bicones of scattered light on either side of
a band of obscuration, possibly associated with an edge-on disk.  A
K-band point source sits within the apex of the brighter, northeastern
cone, while a second, much redder source occupies the southern
cone. Notably, this second source is closest to the faint SCUBA core
FIR~6, although again these features are not precisely coincident.

There are additional cores detected at 850\mic\ that
are not associated with bright 8.0\mic\ sources (Table~\ref{scuba}).
These may harbour very young sources or be pre-stellar cores.
North of DR21(OH) and ERO~3, it is interesting to note that the
850\mic\ emission FIR~1/FIR~2/FIR~3 roughly tracks the chain of
reddened stars rather than the dark patch where very few stars are
seen (even with Spitzer).  Even so, it seems clear that the 8.0\mic\
peaks and SCUBA cores are associated with different populations of
young objects. The SCUBA cores may eventually evolve into compact
clusters of young stars, much like those seen with Spitzer and WFCAM
(the ``IRS'' sources), in which a dominant 8\mic\ peak, probably an
early B-type star (Paper II), is surrounded by a number of fainter,
lower-mass sources.


 \begin{table*} \begin{minipage}{175mm} \caption{H$_2$ knot or outflow
 parameters (excluding the main DR21 and W75N flows)}
 \begin{tabular}{@{}lccccccl@{}} \hline Jet & RA$^a$ & Dec$^a$ & l$^b$
 & P.A.$^c$ & $\theta ^d$ & $F_{\rm 1-0S(1)} ^e$ & Descriptive note \\
 & (2000.0) & (2000.0) & (arcsec) & (degs) & (degs) &
 ($10^{-18}$W~m$^{-2}$) \\ \hline

A~1-1:A~1-3  & 20:39:19.2 & 42:14:53 & 71  & 169       & 3--10 & 20  & Collimated jet \\
A~2-1:A~2-4  & 20:39:17.0 & 42:16:14 & 27  & $\sim$45  &  60   & 200 & Compact, bipolar jet from ERO~1 \\
A~3-1        & 20:39:01.9 & 42:18:11 & --  & --        &  --   & 38  & Arc of H$_2$ emission \\
A~4-1:A~4-2  & 20:38:58.9 & 42:18:13 & 48  & 24        & 3--29 & 20  & Possible faint jet and bow \\
A~5-1:A~5-3  & 20:38:56.4 & 42:18:16 & 55  & 29        & 5--30 & 58  & Collimated jet \\
A~6-1:A~6-2  & 20:38:41.3 & 42:19:01 & --  & --        & --    & 60  & Possible bow shock \\
A~7-1        & 20:38:46.1 & 42:19:14 & --  & --        & --    & 45  & Bright, extended knot \\
A~8-1        & 20:38:47.5 & 42:19:22 & --  & --        & --    & 26  & Bright, diffuse knot \\
A~9-1:A~9-3  & 20:38:50.5 & 42:19:09 & 35  & 49        & 10--30& 56  & Possible collimated jet \\
A~10-1       & 20:39:03.7 & 42:20:15 & 12  & 47        &  40   & 35  & Collimated jet \\ \\

B~1-1:B~1-3  & 20:38:53.1 & 42:20:08 & $\sim$45 & 111  &  45   & 43  & Jet and bow shock \\
B~2-1        & 20:38:55.0 & 42:20:39 & --  & --        &  --   &$<$10& Faint, elongated knot \\
B~3-1:B~3-2  & 20:38:59.9 & 42:20:51 & --  & --        &  --   & 15  & Two or more fingers of emission \\
B~4-1        & 20:38:58.3 & 42:21:09 & 12  & 175       &  --   & 39  & Extended bow or ``bullet'' \\
B~5-1        & 20:39:04.5 & 42:22:44 & --  & --        &  --   &$<$10& Possible extension of B6-1:B~6-3 \\
B~6-1:B~6-3  & 20:39:00.2 & 42:23:02 & 68  & 104       & 6--11 & 172 & Bright, knotty jet \\
B~7-1:B~7-3  & 20:38:58.0 & 42:22:51 & 32  & $\sim$24  &  12   & 14  & Curving jet \\
B~8-1        & 20:38:53.4 & 42:23:27 & --  & --        &  --   &$<$10& Pair of faint knots \\
B~9-1        & 20:39:04.6 & 42:26:06 & 28  & $\sim$88  &  50   & 124 & Series of bright knots \\
B~10-1       & 20:39:07.1 & 42:26:20 & --  & --        &  --   &$<$10& Faint, elongated knot \\
B~11-1:B~11-2& 20:38:45.4 & 42:24:58 & 30  & 150       &   9   & 12  & Possible faint/diffuse jet \\
B~12-1:B~12-2& 20:38:45.9 & 42:24:16 & 20  & 64        &  --   & 22  & Bright knot in possible faint jet \\ 
B~13-1       & 20:38:55.3 & 42:21:27 & --  & --        &  --   &$<$10& Faint, curved knot (bow shock?) \\
B~14-1:B~14-3& 20:39:17.3 & 42:23:46 &100  & 99        &   7   &$<$10& Long chain of 3 faint knots/bows \\ 
B~15-1       & 20:39:03.9 & 42:24:58 & --  & --        &  --   &$<$10& Faint, elongated knot \\
B~16-1       & 20:39:09.6 & 42:25:24 & --  & --        &  --   &$<$10& Faint, elongated knot \\ \\

C~4-1        & 20:38:38.0 & 42:38:15 & --  & $\sim$61  &  --   &$<$10& Possible elongated jet \\
C~5-1:C~5-2  & 20:38:53.0 & 42:37:47 & --  & --        &  --   &$<$10& Faint jet \\
C~7-1        & 20:38:50.6 & 42:39:09 & --  & --        &  --   & 50  & Bright, curved knot (bow shock?) \\
C~8-1:B~8-5  & 20:38:30.1 & 42:39:21 & 80  & 57        & 5--10 & 54  & Collimated jet \\
C~9-1:B~9-2  & 20:38:30.7 & 42:39:52 & 24  & 121       &   8   &  8  & Possible jet \\ 
C~10-1:B~10-2& 20:38:20.1 & 42:39:12 & --  & --        &  --   &$<$10& Two faint knots \\ \\

D~1-1:D~1-8  & 20:37:44.8 & 42:37:12 & 182 & $\sim$18  &  18   & 190 & Bipolar outflow with bow shocks \\
D~2-1        & 20:38:00.8 & 42:40:13 & --  & --        &  --   &$<$10& Faint arc of emission \\
D~3-1:D~3-2  & 20:38:00.4 & 42:41:32 &$\sim$25&$\sim$90&  --   & 44& Two extended knots or ``bullets''\\
D~4-1:D~4-6  & 20:37:48.3 & 42:43:42 &  76 & 99        &   7   & 58  & Collimated jet \\ 
D~5-1:D~5-3  & 20:38:09.7 & 42:38:07 &  35 & 56        &  --   &$<$10& Faint chain of knots \\
D~6-1:D~6-3  & 20:38:05.6 & 42:37:42 &  28 &  6        &  --   &$<$10& Faint chain of knots \\
D~7-1:D~7-2  & 20:38:04.8 & 42:39:05 & --  & --        &  --   &$<$10& Two faint knots (possible jet) \\
D~8-1        & 20:38:02.2 & 42:39:12 & --  & --        &  --   &$<$10& Faint arc (bow shock?) \\
D~9-1        & 20:38:00.6 & 42:39:35 & --  & --        &  --   &$<$10& Faint arc (bow shock?) \\
D~10-1       & 20:38:03.2 & 42:39:49 & --  & --        &  --   &$<$10& Faint, compact knot \\
D~11-1       & 20:37:52.3 & 42:40:50 & --  & --        &  --   &$<$10& Faint knots and filaments \\
D~12-1:D~12-3& 20:37:44.7 & 42:40:27 &$\sim$50&$\sim$110 &--   &$<$10& Faint group of knots (possible jet) \\ \\

E~1-1        & 20:36:59.8 & 42:11:19 & 24  & 118       & 4--11 &$<$10& Faint knotty jet \\
E~2-1:E~2-3  & 20:37:02.3 & 42:11:51 & 21  & $\sim$139 &   18  & 23  & Possible knotty jet \\
E~3-1:E~3-4  & 20:37:00.1 & 42:13:14 &$\sim$50&$\sim$92&   40  & 80  & Curving knotty jet \\
E~4-1        & 20:37:21.0 & 42:16:03 &$\sim$40&  124   &   30  & 36  & Knotty jet \\
E~5-1:5-6    & 20:37:28.9 & 42:16:12 & 130 & $\sim$115 &   12  & 64  & Extended, knotty jet \\ 
E~6-1:6-2    & 20:37:19.0 & 42:11:39 & 40  &  95       &  --   &$<$10& Two faint knots \\ 
E~7-1        & 20:37:14.5 & 42:13:01 & --  &  --       &  --   &$<$10& Faint bow shock \\ 
E~8-1        & 20:37:05.2 & 42:13:14 & --  &  --       &  --   &$<$10& Faint knot (part of E~3-1:E~3-4?)\\ 
E~9-1        & 20:37:32.0 & 42:14:27 &$\sim$25&$\sim$175& --   &$<$10& Faint north-south filament \\ 

 \hline
 \label{jets}
 \end{tabular}
 $^a$Coordinates of the first H$_2$ feature in each jet, A~1-1, A~2-1, B~1-1, etc. 
     (see Figs.~\ref{h2jetsa} to \ref{h2jetse}). \\
 $^b$Length of the H$_2$ flow, when multiple components are identified. 
     At 3~kpc, 1\arcsec\ = 0.0145~pc = 3000~AU. \\  
 $^c$Jet position angle (E of N), again, when multiple components are identified. \\
 $^d$Flow opening angle (see text for details). \\
 $^e$Integrated H$_2$ 1-0S(1) flux, uncorrected for extinction, accurate to 20-30\%. 
 \end{minipage}
 \end{table*}

 \begin{table*}
 \centering
 \begin{minipage}{175mm}
  \caption{Photometry of possible H$_2$ jet sources (excluding the main DR21 and W75N flows)}
  \begin{tabular}{@{}llccccccccc@{}}
  \hline
Jet        & Source & RA     & Dec          &  K     & [3.6] & [4.5] & [5.8] & [8.0] & $\alpha$ & Other\\
           & Ref~$^a$&(2000.0)&(2000.0)     &        &       &       &       &       &          & name \\
 \hline

A~2-1:A~2-4  & a & 20:39:16.73 & 42:16:09.2 & 10.98  & 8.12  & 7.06  & 5.50  & 4.82  & 1.1 & ERO~1\\
A~3-1        & b & 20:39:01.48 & 42:18:09.7 & 15.37  & 14.55 & 12.88 & 11.88 & 10.53 & 1.7 & \\
A~4-1:A~4-2  & c & 20:39:01.10 & 42:18:59.0 &   --   & 15.14 & 13.17 & 11.95 & 10.27 & 2.6 & \\
A~5-1:A~5-3  & d & 20:38:57.35 & 42:18:32.9 & 14.67  & 13.85 & 13.11 & 11.29 & 9.60  & 2.2 & \\ \\

B~7-1:B~7-3  & e & 20:38:57.20 & 42:22:41.1 & 8.34   & 7.40  &  6.62 &  5.53 & 4.13  & 1.0 & DR21-IRS~6 \\
B~6-1:B~6-3  & f & 20:39:01.90 & 42:23:02.3 &  --    & 14.82 & 13.57 & 11.87 & 10.26 & 2.5 & \\ 
             & g & 20:38:59.97 & 42:23:07.7 & 19.6   & 15.47 & 13.77 & 12.30 & 10.70 & 2.6 & \\ 
B~9-1        & h & 20:39:04.06 & 42:26:07.8 & 15.22  & 13.22 & 11.36 & 10.46 & 9.04  & 1.8 & \\ \\

C~8-1:B~8-5  & j & 20:38:33.30 & 42:39:38.5 & 15.92  & 12.51 & 10.63 & 9.26  & 8.06  & 2.2 & \\ 
             & k & 20:38:32.88 & 42:39:35.7 &  --    & 13.02 & 11.09 & 10.04 & 9.16  & 1.5 & \\ 
C~9-1:B~9-2  & l & 20:38:32.32 & 42:39:45.2 & 14.18  & 11.19 & 9.70  & 8.76  & 7.80  & 1.0 & \\ \\

D~3-1:D~3-2  & m & 20:38:08.27 & 42:38:36.4 &  --    &  7.09 & 6.26  & 4.80  & 4.26  & 0.6 & W75N-IRS~7\\ \\

E~4-1        & n & 20:37:17.66 & 42:16:37.4 &  --    & 10.62 & 10.03 & 6.83  & 5.04  & 4.1 & L906E-IRS~1\\

 \hline
 \label{sources}
 \end{tabular}
 \smallskip \\
 $^a$Possible source of the jet identified in column 1 (see appendix for details). 
 \end{minipage}
  \end{table*}


\subsection{The \htwo\ outflows and their driving sources}
 
For the sake of brevity we consign our description of the many H$_2$
jets and outflows in DR21/W75 to the appendix.  In Figs.~\ref{h2jetsa}
to \ref{h2jetse} we show continuum-subtracted H$_2$ images of various
regions along and adjacent to the DR21/W75 molecular ridge.  For
simplicity we split the DR21/W75 region into five areas, which we
label A--E (see Fig.\ref{wfcam0}). We then assign numbers to individual
H$_2$ features in each region.  Knots that appear to be part of the
same jet are given consecutive numbers.  For example, knots A~1-1,
A~1-2 and A~1-3 are possibly part of the same jet in area A, while
A~2-1, although still in area A, is probably part of a different flow.

In Table~\ref{jets}, where possible, jet lengths, position angles and
jet opening angles are estimated, although many emission-line features
are single knots with no obvious central engine.  We do not discuss
the main DR21 outflow, since this H$_2$ flow has been studied in
considerable detail in past papers \citep[e.g.][]{gar91a,dav96,smi98}.
In all we identify about 50 separate outflows.

Although in previous sections we have identified and labeled the
bright K-band/8.0\mic\ sources, these are not necessarily the outflow
sources.  We have therefore used the Spitzer IRAC images to identify
young stellar objects (YSOs) based on their spectral index.
As mentioned earlier, $\alpha$ is derived from linear fits to the
observed flux density in each of the four IRAC bands.  Note
that the more nebulous sources, like DR21-IRS~1 and W75N-IRS~1, were
not retrieved by our analysis technique (described in Paper II) and
therefore have not been flagged as YSOs.
 
Traditionally, YSOs have been classified as Class 0, I, II or Class III
based on their bolometric temperature, infrared spectral slope
(measured between 2\mic\ and 20\mic ) and the associated colour excess
due to a dusty envelope \citep{lad87,and93}.  Class 0 SEDs consist
entirely of grey-body emission from the cool circumstellar envelope
and are thus detectable only at mid- to far-infrared wavelengths
\citep[e.g.][]{nor04}. Class I, flat-spectrum, Class II and Class III 
YSOs, detectable in the near-infrared, have photospheric SEDs with an
infrared excess, the slope of which, $\alpha$, is positive for
the younger (Class I) sources.  
More recently, \citet{lad06} have
applied this classification scheme to Spitzer photometry of YSOs in
the low-mass star forming region IC~348.  They distinguish stars with
optically thick circumstellar disks from their ``anemic''
counterparts, sources with heavily depleted, optically-thin disks or
disks with large central holes.  Their thick-disk sources have SEDs
measured from the IRAC bands of $\alpha > -1.80$, while the anemic
disk sources have shallower spectral slopes with $-1.80 > \alpha >
-2.65$.

In principle, IRAC colour-colour diagrams can also be used to
distinguish YSO classes \citep{meg04,all04,lad06}.  In the four IRAC
bands centred at 3.6\mic , 4.5\mic , 5.8\mic\ and 8.0\mic , Class 0
and Class I protostars have roughly the same colours;
[3.6]-[4.5]$>$0.7 and [5.8]-[8.0]$>$1.0, indicative of mass accretion.
Class II sources (T~Tauri stars) overlap this region, roughly occupying
a box with 0.0$<$[3.6]-[4.5]$<$0.8 and 0.3$<$[5.8]-[8.0]$<$1.1.
Reddened Class II YSOs have somewhat higher [3.6]-[4.5] colours,
although slightly lower [5.8]-[8.0] colours.  However, as we discuss
in Paper II, there is considerable overlap in colour-colour space
between the various YSO classes.
\citet{meg04} note that from Spitzer colours alone it is possible to
mis-identify reddened Class II sources or Class II sources with edge-on
disks as Class I sources.  Also, strong silicate absorption at
9.7\mic\ and water-ice absorption features at 3\mic\ and (to a lesser
extent) 6\mic\ may modify the Spitzer colours of massive young stars,
increasing [3.6]-[4.5] though reducing [5.8]-[8.0] \citep{gib00,
dis04}. So although Spitzer remains a powerful tool for distinguishing
YSOs from background and/or foreground field stars, particularly in
high-extinction regions where near-IR data are of limited use, here we
only identify sources with reddened photospheres, and distinguish
protostars ($\alpha > 0.25$) from pre-main-sequence objects ($0.25 >
\alpha > -1.60$).

In Figs.~\ref{h2jetsa} to \ref{h2jetse} we plot the positions of all
of the YSOs that were extracted from the Spitzer images {\em in all
four IRAC bands}, as reported in Paper II. We use circles to mark the
protostars and triangles to indicate the remaining reddened
photospheres.  Approximately 1600 sources were retrieved across the
entire DR21/W75 field of which 165 were protostars and an additional
482 were pre-main-sequence objects (we have only identified point
sources with photometric errors of less than 0.05 mag at 8.0\mic ). If
we consider that some of the protostars will actually be Class II YSOs
viewed edge-on, then the YSO population is reasonably consistent with
an order-of-magnitude increase in the duration of the Class II phase
over the Class 0/I protostellar phase. The large-scale distribution
and clustering of the YSOs in DR21/W75 is discussed in Paper II.

The K-band magnitudes, mid-IR magnitudes and spectral indices of the
candidate outflow sources are given in Table~\ref{sources}.  Most are
undetected at the shorter near-IR wavelengths; three were undetected
even at K (fainter than $\sim$19).  Two of the brighter targets were
saturated in our K-band data.  The sources are assigned a letter in
Table~\ref{sources} and in each of the Figures in the appendix (note
that some sources have both an IRS designation and an outflow source
reference letter).


\section{Discussion}

\subsection{Properties of the H$_2$ jet sources}
 
\begin{figure}
     \epsfxsize=8.0cm
\epsfbox{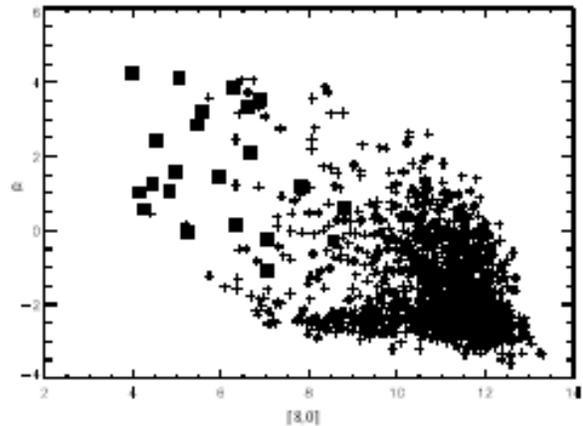}
 \caption[]{Plot of spectral index, $\alpha$, against source magnitude at 
 8.0\mic .  The sources listed in Table~\ref{irs} are marked with 
 squares, while the remaining IRAC sources are marked with crosses.  }
  \label{alpha}
  \end{figure}

In Fig.~\ref{alpha} we plot mid-IR spectral index, $\alpha$, against source
magnitude at 8.0\mic\ for all point sources in the DR21/W75 field
detected in all four IRAC bands.  The plot suggests a mild correlation
between source magnitude and spectral index.  In other words, many of
the bright 8.0\mic\ sources appear to be embedded.  The brighter
8.0\mic\ sources are therefore probably luminous young stars.
Accurate luminosity estimates for individual sources are difficult to
measure without photometry at longer wavelengths: note that the
brightest sources were saturated in the MIPS images of this region
\citep{mar04}, and that IRAS data lack sufficient spatial resolution to
distinguish individual stars.  Even so, it seems safe to assume that
the intermediate and high-mass sources correspond with the reddest
8.0\mic\ sources in Table~\ref{irs}.  

We find little evidence for H$_2$ jets associated with the 8.0\mic\
``IRS'' sources (see the Appendix for a detailed discussion), and only
one of the four EROs has a definite H$_2$ jet. Most of the candidate
jet sources in Table~\ref{sources} are relatively faint at 8.0\mic ,
while $\sim$80\% of the outflows listed in Table~\ref{jets} have no obvious
source identified at 8.0\mic . It thus seems clear that, with the
exception of the massive DR21 and W75N outflows themselves, the H$_2$
jets are not driven by the massive protostars.  Instead, low- or at
most intermediate-mass YSOs seem to power each flow.

The candidate outflow sources are, however, very red (in comparison to
the criteria used to select protostars from the Spitzer observations).
Many will probably be Class 0 or early Class I sources.  Some may be
flat-spectrum or Class II YSOs viewed edge-on and therefore through
their circumstellar disks, although if this were the case one would
expect to see more Class II jet sources (with negative values of
$\alpha$) viewed pole-on.  It seems likely that many of the H$_2$
flows are powered by low-mass ``protostars'' that are too faint and/or
too heavily embedded to be detected even by Spitzer.

Candidate central engines for the massive DR21 and W75N outflows
remain elusive.  Accurate Spitzer photometry of the IRS sources in
each region is difficult, since at near- and mid-IR wavelengths these
sources are often extended and bathed in diffuse emission.  The
brighter sources are also saturated at the longer wavelengths.  In
DR21, \citet{hsm05} note that UCH{\sc ii} region DR21(D) is detected
in the Spitzer bands and has an extremely red mid-IR SED.  Even so, it
is not well placed mid-way between the extensive H$_2$ flow lobes.
Also, it does not coincide with the brightest dust-continuum peak in
our 850\mic\ data.  DR21(OH), on the other hand, {\em is} associated
with a SCUBA core; it is also detected at mid-IR wavelengths, and
again is extremely red.  Its outflow is clearly traced in methanol
maser emission \citep[][see also Fig.~\ref{h2jetsb}]{pla90,kog98},
though DR21(OH) itself is associated with multiple water masers
\citep{for78,man92} and therefore probably comprises multiple sources,
any one of which could drive the outflow.

The relationship between the infrared sources and multiple UCH{\sc ii}
regions in W75N are discussed in detail by \citet{she04}.  The mid-IR
properties of these sources are analysed by \citet{per06}.  From their
analysis of ground-based and Spitzer mid-IR photometry
they propose that W75N-IRS~2 is a B3 star, while W75N-IRS~3 and
W75N-IRS~4 are intermediate-mass Class II YSOs.  In Fig.~\ref{spit3}
W74N-IRS~1 is clearly elongated; here \citet{per06} identify two sources,
one coincident with the near-IR W75N-IRS~1 peak, and a second, much
redder source coincident with the MM1 dust continuum peak
\citep{she03} and VLA radio sources \citep{tor97}.  
This secondary mid-IR source is almost certainly associated with the
W75N outflow source.  Its position, RA(2000):20$^h$38$^m$36.5$^s$
DEC(2000):42\dg 37\arcmin 33.6\arcsec is also notably coincident, to
within an arcsecond, with the ``W75N'' 850\mic\ peak in
Table.~\ref{scuba}.

Overall, W75N and DR21 are similar in many ways.  Both regions harbour
a massive core containing clusters of H{\sc ii} regions, some of which
are ultra-compact.  Both contain water, OH and methanol masers.  Both
are also associated with bright, near- and mid-IR nebulosity and
clusters of tens of embedded sources, many of which are B-type
(5--10~\Msolar ) stars.  Both are also surrounded by smaller, more
compact clusters of just a few (or just one) early B-type stars plus
associated low-mass stars.  DR21 and W75N themselves drive very
massive molecular outflows, which are almost parallel on the sky (as
is the flow from DR21(OH)).  And finally, numerous smaller-scale H$_2$
jets and outflows are found in the vicinity of each massive
star-forming region, these jets being driven by low or at most
intermediate-mass young stars.


\subsection{General outflow characteristics: sizes, opening angles and 
orientations}

Estimated jet parameters for the more prominent outflows are given in
Table~\ref{jets}.  At a distance of 3~kpc a parsec-scale jet should be
$>$70\arcsec\ in length.  At least five of the H$_2$ outflows
meet this requirement.  For a modest flow velocity of 100\kms\ a jet
should reach a length in excess of a parsec in $\sim 10^4$~yrs; thus,
a Class 0 source could easily be associated with a parsec-scale flow.
Some of the long jets in Table~\ref{jets}, like B~6-1:B~6-3 and
C~8-1:C~8-5 do have candidate sources with SED slopes consistent with
very red YSOs (Table~\ref{sources}).


Precise jet opening angles ($\theta$) are rather difficult to measure,
particularly at kilo-parsec distances.  The emitting flanks of
large-scale bow shocks are often much wider than the underlying jet
\citep[HH~212 is a spectacular example;][]{zin96}, while changes in 
flow direction over time will increase the apparent flow opening
angle.  Moreover, if one includes all features along the flow axis, or
just the extent of the leading bow shock, very different values can be
inferred. Consequently, for a few flows in Table~\ref{jets} we list a
range of values; the smaller is measured between the candidate outflow
source and the knot furthest from the source, while the larger value
represents the maximum apparent flow opening angle that includes all
emission features.  If no YSO is identified, the angles are measured
between the two most distant knots in the jet.  Overall, the jets in
the DR21 and W75N regions appear to have high degrees of collimation,
consistent with the apparent youth of the candidate central engines in
Table~\ref{sources}.

Flow position angles on the sky are also listed in Table~\ref{jets}.
The two main DR21 and W75N outflows have almost the same position
angle ($\sim$60\dg -70\dg ).  The DR21(OH) molecular outflow, which is not
obviously traced in H$_2$, has a similar ``east-west'' position angle
\citep{lai03}, as does the prominent H$_2$ jet 
B~6-1 to B~6-3.  Each of these flows is orientated roughly orthogonal
to the chain of 850\mic\ cores and the molecular ridge that runs
north-south through DR21 towards W75N.  If we simply count the number
of flows in regions A and B (including the main DR21, DR21(OH) and
W75N outflows) with position angles in the range 45\dg --135\dg\, and
compare this to the number of flows with angles in the range 0\dg
--45\dg\ and 135\dg --180\dg , we find a ratio of 11 to 6. (For the
whole region imaged, the ratio is 25:10.)  Moreover, a
Kolmogorov-Smirnov test yields a probability of only 53\% that the
distribution of angles in areas A and B is homogeneously distributed
between 0\dg\ and 180\dg .  It is therefore possible that the flow
position angles {\em are} to some degree dictated by the large-scale
cloud morphology.  If the north-south elongation of the DR21/W75 ridge
is produced by gravitational collapse along east-west orientated
magnetic fields lines, then the random motions of individual star
forming cores within the ridge are only moderately successful at
scrambling the polar axes of the outflow sources.  This process may be
even less effective with the very massive YSOs, DR21, W75N, ERO~1 and
DR21(OH), where the flows clearly are orthogonal to the
north-south molecular ridge that extends through DR21.  Note that
\citet{val06} find that the magnetic field direction, inferred from
850\mic\ linear polarimetry observations, is orientated roughly
orthogonal to the DR21 filament, as expected.


\subsection{Spatial distribution of outflows and their relationship to 
YSO clustering}

In busy star forming regions like DR21 and W75N, interactions between
neighbouring stars as they pass by each other may inhibit disk
formation, or even strip stars of their circumstellar disks.  Since
accretion drives outflows, YSOs without actively accreting disks will
obviously not produce H$_2$ jets.  In Figs.~\ref{h2jetsa} to
\ref{h2jetse} one can see that clusters of YSOs do not generally coincide
with an abundance of H$_2$ features.  Although these areas are not
entirely devoid of H$_2$ jets, H$_2$ features do seem to be more
abundant in other regions, which show a paucity of YSOs (as identified
by our Spitzer analysis), e.g. the region $\sim$1\arcmin\ northwest of
the western DR21 outflow lobe (Figs.~\ref{h2jetsa} and \ref{h2jetsb})
and 3\arcmin --5\arcmin\ west of the busy L906E-IRS~3/IRS~4/IRS~5
cluster (Figs.~\ref{h2jetse}). A stellar density plot is shown in
Paper II.

To investigate this issue further, we consider whether outflow sources
have fewer neighbours than YSOs with no obvious H$_2$ jet. To do this
we count the number of neighbours each YSO has within a radius of 50
pixels (20\arcsec , or 0.3\,pc). YSO neighbours are classified as
Spitzer-identified protostars ($\alpha > 0.25$) {\em or}
pre-main-sequence stars ($0.25 > \alpha > -1.60$), since all of these
sources are likely to be in the same star forming regions and
therefore potentially involved in star-star interactions or disk
stripping.  However, because all candidate outflow sources appear to
possess reddened photospheres (e.g. Table~\ref{sources}), we count
neighbours only for the protostars.  We have confined the analysis to
a 8.0\arcmin $\times$10.7\arcmin\ (1200$\times$1600 pixel) region
centred near DR21-IRS~5, a 8.0\arcmin $\times$6.7\arcmin\ region
centred on W75N, and a 8.0\arcmin $\times$6.7\arcmin\ region centred
midway between L906E-IRS~3 and L906E-IRS~6. In other words, we focus
only on regions where H$_2$ outflows are abundant (except for the
region west of W75N, where we lack complete Spitzer data and therefore
source classifications).

\begin{figure}
     \epsfxsize=8.0cm
\epsfbox{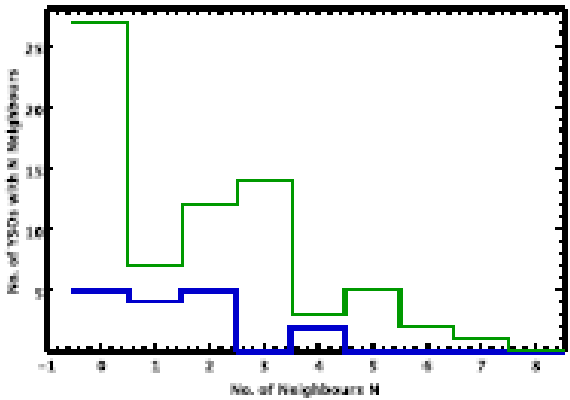}
 \caption[] {Plot of the number of YSO neighbours 
  within a 50-pixel (20\arcsec ) radius of protostars that are  thought
  to drive outflows (thick/blue line) and those that have no obvious H$_2$ jets 
  (thin/green line).}
 \label{hist}
  \end{figure}

The resulting histograms are plotted in Fig.~\ref{hist}.  The mean
number of neighbours for candidate outflow sources is 1.4, while the
mean for the remaining protostars is 1.8.  Both
values are, however, skewed by the large number of sources
with no neighbours (the N$=$0 column in Fig.~\ref{hist}).  Many of
these sources lie on the periphery of the north-south DR~21 ridge, and
around the edges of the W75N and L906E cores.  If we exclude the
sources with no neighbours from the statistics, then the outflow
sources have a mean of 2.0 neighbours, while the remaining YSOs have a
slightly higher mean of 2.9 neighbours. The latter is evident as a
secondary peak in Fig.~\ref{hist}.  The slight increase in the
number of neighbours associated with non-outflow YSOs is an extremely
tentative result, not only because of the modest number of sources
considered, but also because of uncertainties in identifying outflow
source candidates.

At first sight it seems that very few YSOs in the region drive {\em
detectable} H$_2$ flows.  By assuming that individual H$_2$ knots are
parts of extended flows, we identify only 53 H$_2$ outflows in
Table~\ref{jets}, while our analysis of the Spitzer photometry reveals
165 protostars spread over the entire 0.8\dg\ field.  Some of
the Class 0/I protostars will actually be pre-main-sequence Class II
sources seen edge-on, and we do not include some of the very faintest
knots and wisps of emission in Table~\ref{jets}. Moreover, 
(1) the distance to DR~21 will limit our
ability to detect some of the ``weaker'' flows, (2) extinction
\citep[approaching 100-200 magnitudes at V in the densest
cores;][]{moo91b,cha93a,per06} will hide some H$_2$ features, and (3)
east-west orientated outflows will probably break out of the denser
cloud regions and therefore not be particularly bright in H$_2$
emission.  Even so, we conclude that we have detected -- within a
factor of a few -- roughly the same number of H$_2$ outflows as
embedded protostars in DR21/W75.


\subsection{Momentum injection and cloud support}

Supersonic turbulent energy in giant molecular clouds is expected to
decay on timescales of a few million years \citep{ost01}. However,
clouds last longer than this, harbouring multiple epochs of star
formation at any one time. Outflows may be responsible for the
continued injection of turbulent energy in to molecular clouds,
maintaining virial equilibrium and thereby limiting cloud collapse and
the overall star-formation efficiency.

The turbulent momentum (and energy) in a cloud is approximately equal to
the product of the cloud mass and turbulent line width (and line width
squared). The mass of the cores associated with DR21/W75 have been
estimated from modeling of dust continuum observations
\citep{moo91a,cha93a}, molecular line maps
\citep{dic78,wil90} and, more recently, as virial masses from
extensive CS observations \citep{shi03}.  Core masses for DR21,
DR21(OH) and W75N of between $1\times10^3$~\Msolar\ and
$3\times10^4$~\Msolar\ are predicted.
If we adopt a turbulent velocity of $\sim$3--5\kms , estimated from
sub-arminute-resolution maps in gas tracers that are optically thin
and/or probes of high-density cores, like CS, NH$_3$ and C$^{18}$O
\citep[e.g.][]{wil90,cha93b,val06} -- observations that are likely to
yield velocities relatively unaffected by outflows or large-scale
cloud motions -- then a total momentum in the range 
$3\times10^3$ -- $1.5\times10^5$~\Msolar\ \kms\ 
and turbulent energy in the range
$1.8\times10^{40}$ -- $1.5\times10^{42}$~J 
is inferred for each region.

The momentum and energy supplied by an outflow from a young star vary
with time \citep[e.g.][]{smi00}.  Younger flows (from Class 0
sources) are more powerful than their Class I and Class II
counterparts, and therefore provide a larger momentum flux, $P$, and
energy flux (or mechanical power), $L$.  \citet{wal05} note that one
could use the canonical values for $P$ for Class 0 and Class I {\em
solar mass stars} measured by \citet{bon96} to estimate the momentum
supplied over the accreting lifetime of a young star. If $P({\rm Class~0}) 
\sim 6\times10^{-5}$~\Msolar \kms\ yr$^{-1}$ and $P({\rm Class~1})
\sim 4\times10^{-6}$~\Msolar \kms\ yr$^{-1}$, then assuming durations
of $10^4$ and $10^5$ years for the Class 0 and Class I stages, over
its lifetime an outflow from a low-mass protostar should inject
$\sim$1.0~\Msolar \kms\ of momentum into its surroundings.  The
outflow kinetic energy scales with the velocity; since most of the
molecular material in outflows is probably entrained in the relatively
slow-moving wings of bow shocks (which explains why CO outflows are
usually an order of magnitude slower than HH jets), then for a
molecular flow velocity of $\sim$20\kms , the kinetic energy supplied
by a typical outflow would be $\sim4\times10^{37}$~J .  The 50 or so flows
from low-mass stars in DR21/W75 could therefore provide of the order
of $\sim$50~\Msolar \kms\ of turbulent momentum and inject about
2$\times10^{39}$~J of kinetic energy into the cloud over their lifetimes.
Both values are less, by two to three orders of
magnitude, than what is apparently needed to maintain the observed
turbulent momentum and energy in the region.  Note
also that the trend toward east-west orientated outflows noted earlier, and
the small filling factor of each collimated jet (with respect to the
total volume of the cloud) limits the efficiency of momentum and
energy transport to the global GMC environment.

But we have not considered the high-mass sources.  Although these are
relatively few in number, their outflows are more massive and more
energetic, by three to four orders of magnitude.  The momentum in the
DR21 and W75N outflows, measured from CO observations, is of the order
of $\sim 2\times10^4$~\Msolar \kms\ and $10^3 - 10^4$~\Msolar \kms ,
respectively \citep{gar91b,dav98a,she01}.  This would be more than
enough to provide turbulent support against cloud collapse, if this
momentum was distributed throughout the cloud.  But of course these
two flows do not impinge on the whole cloud; both are fairly well
collimated and so probably efficiently ``drill'' through their
molecular surroundings, rather than feed turbulence into the larger
environment.  Note that massive dense cores, traced at 850\mic ,
remain around DR21, DR21(OH) and W75N (Fig.~\ref{scu}); they are not
entirely dispersed by the powerful outflows.
 
Finally, rather than use ``mean'' values for the momentum and energy of
the outflows, we can infer these directly from the integrated H$_2$
line fluxes, since the outflow mechanical power should be directly
related to the energy radiated in the entraining shock fronts.
\citet{dav96} point out that in strong-shock prompt entrainment
scenarios, where the ambient gas is swept up in a momentum-conserving
fashion to form the molecular ``CO'' outflow, the energy radiated in
the wings of the entraining bow shock
-- the flux seen in H$_2$ and other coolants -- should be equivalent
to the mechanical power in the swept-up outflow.  In regions A, B and
C, the total H$_2$ 1-0S(1) line flux, $F_{\rm 1-0(S1)}$, measured from
the H$_2$ flows identified in Table~\ref{jets} (again not including
the special case DR21 and W75N flows) is about
$1\times10^{-15}$~W~m$^{-2}$.  At a distance $d\sim$3~kpc, and
assuming that the 1-0S(1) line represents 10\% of all H$_2$ line
radiation, then the total H$_2$ luminosity will be $L_{\rm H2} = 10
\times L_{\rm 1-0(S1)} \times 4 \pi d^2 = 1\times10^{27}$~W.  
Since there will be other important molecular coolants; H$_2$O, CO,
etc. \citep{ben00,gin02}, the energy radiated in the outflow bow
shocks will actually be 2-3 times higher.  The observed line fluxes
are also not corrected for extinction.  The observed jets probably
occupy regions of fairly modest extinction; an $A_v$ of 10 would boost
the luminosity by a further factor of 2.5.  The total radiated power
is therefore probably an order of magnitude higher, $\sim10^{28}$~W.
For a dynamical age of $\sim3\times10^{11}$~sec (1~pc/100\kms ) the
energy pumped into the cloud equates to $\sim3\times10^{39}$~J, which
again seems to be insufficient to account for the turbulent energy in
the GMC.

In conclusion, the {\em current} generation of low-mass YSO flows
probably does not drive the observed cloud turbulence.  However, many
repeats of the current epoch of outflow activity, possible during the
few million year timescale for turbulent decay in the GMC, might
provide sufficient momentum and energy. Notably, the $10^3 -
10^4$~year outflow dynamical timescale is a factor of 10-100 times
less than the turbulent decay time scale (for a $\sim$3~pc cloud and
velocity of $\sim$3~\kms ), as required.  Imaging surveys of other
high-mass star forming regions are required to establish whether the
abundance of outflows in DR21/W75 is common in other regions, or
whether DR21/W75 is uniquely active.  The former would support the
occurance of repeat epochs of outflow activity in this and other
regions.


\section{Summary and Conclusions}

We compare near-IR and mid-IR images of the high-mass star forming
regions DR21 and W75, obtained with the wide-field camera, WFCAM on
the U.K. Infrared telescope, and the IRAC camera on the Spitzer Space
telescope.  850\mic\ images of much of the region, obtained with the
SCUBA bolometer array on the James Clerk Maxwell Telescope, are also
considered.

The H$_2$ 1-0S(1) images obtained with WFCAM reveal numerous
line-emission features, most of which appear to be associated with
collimated flows.  The Spitzer IRAC images were found to be of limited
use in tracing these outflows, because of the intense PAH emission in
the IRAC bands associated with background nebulosity, and because of
limited spatial resolution in these very busy star-forming
regions. However, the IRAC data were extremely useful for detecting
YSOs, particularly massive, embedded sources, and for identifying
potential outflow central engines.  The near- and mid-IR photometry
and stellar population are discussed in Paper II. The conclusions
reached in this paper (Paper I) are described below.

\begin{enumerate}

\item The well-known W75N and DR21 high-mass star forming regions are
      surrounded by smaller clusters that harbour perhaps just one
      intermediate or high-mass star (the DR21-IRS and W75N-IRS
      sources). In the case of DR21, the SCUBA data reveal less
      evolved cores that are not generally associated with the
      ``IRS-type'' clusters, although these cores may eventually
      evolve into such clusters. The fact that the SCUBA cores are
      already distributed along the DR21 ridge suggests that the cores
      are dynamically distributed throughout the region {\em before}
      they evolve into intermediate or high-mass star forming
      clusters. As noted in Paper II, the more evolved YSO population
      traced by Spitzer is even more widely distributed.
    
\item Although we find a large number of H$_2$ emission-line features,
      many of which appear to be associated with collimated jets,
      these are usually not associated with either the young SCUBA
      cores or the compact infrared clusters noted above.  Instead,
      they seem to be driven by independent protostars, probably
      low-mass Class 0/I sources distributed around the periphery of
      the dense 850\mic\ cores and north-south molecular ridge.

\item H$_2$ features associated with at least 50 individual outflows
      are identified.  Many are knotty and well collimated, much like
      jets in nearby low mass star forming regions.  At least five are
      parsec-scale flows.  The orientations of the flows on the sky
      show some degree of order, with two thirds of the flows being
      roughly perpendicular (within a 90\dg\ cone) to the north-south
      molecular ridge; we find only a 53\% chance that the flows are
      randomly orientated.  

\item We find very tentative evidence that clustering may inhibit disk
      accretion and thus the production of extensive outflows.  We
      certainly do not see an enhancement in the number of outflows
      around compact clusters or groups of embedded protostars.

\item Although the low-mass YSO flows are abundant and widely distributed
      throughout the region, the observed outflow activity needs to be
      repeated 10-100 times if the flows are to provide sufficient
      momentum and kinetic energy to account for the turbulent motions
      in the large-scale molecular cloud. The current high-mass YSO
      flows associated with DR21 and W75N {\em are} sufficiently
      energetic.  However, because these massive flows are well
      collimated, orientated orthogonal to the clumpy GMC that runs
      through DR21/W75, and because they are few in numbers, the
      transport of their momentum and energy to the ambient medium is
      probably relatively inefficient.

\end{enumerate}

The mid-IR and far-IR observations very effectively trace the
distribution of protostars and pre-stellar cores across the entire
region.  The narrow-band H$_2$ images complement these data by showing
regions that are dynamically active, areas that may otherwise go
unnoticed.  Together, the UKIRT/WFCAM, Spitzer/IRAC and JCMT/SCUBA
data yield a near-complete picture of star formation across the
extensive region targeted here.  Similar wide-field studies in the
future at modest to high spatial resolutions -- particularly with the
advent of SCUBA-2 -- should continue to improve our understanding of
both low and high-mass star formation on global scales.


\section*{Acknowledgments}

We would like to thank the team at CASU for processing the near-IR
data, and the WFCAM Science Archive in Edinburgh for making the data
available to us. We also thank A. Scholz for providing us with a
program to perform a Kolmogorov-Smirnov test, and A. Gibb for sharing
his SCUBA observations of DR21 prior to publication. We also
acknowledge the referee for his speedy review of both Paper I and
II. The United Kingdom Infrared Telescope is operated by the Joint
Astronomy Centre on behalf of the U.K. Particle Physics and Astronomy
Research Council.  The WFCAM data reported here were obtained as part
of the UKIRT Service Programme.  his work was supported by a grant
POCTI/1999/FIS/34549 and POCTI/CFE-AST/55691/2004 approved by FCT and
POCTI, with funds from the European community programme
FEDER. D. Froebrich received funding from the CosmoGrid project,
funded by the Program for Research in Third Level Institutions under
the National Development Plan and with assistance from the European
Regional Development Fund.  This research made use of data products
from the Spitzer Space Telescope Archive. These data products are
provided by the services of the Infrared Science Archive operated by
the Infrared Processing and Analysis Center/California Institute of
Technology, funded by the National Aeronautics and Space
Administration and the National Science Foundation.


\appendix

\section[]{Description of the H$_2$ jets}

\begin{figure*}
     \epsfxsize=16.0cm
     \epsfbox{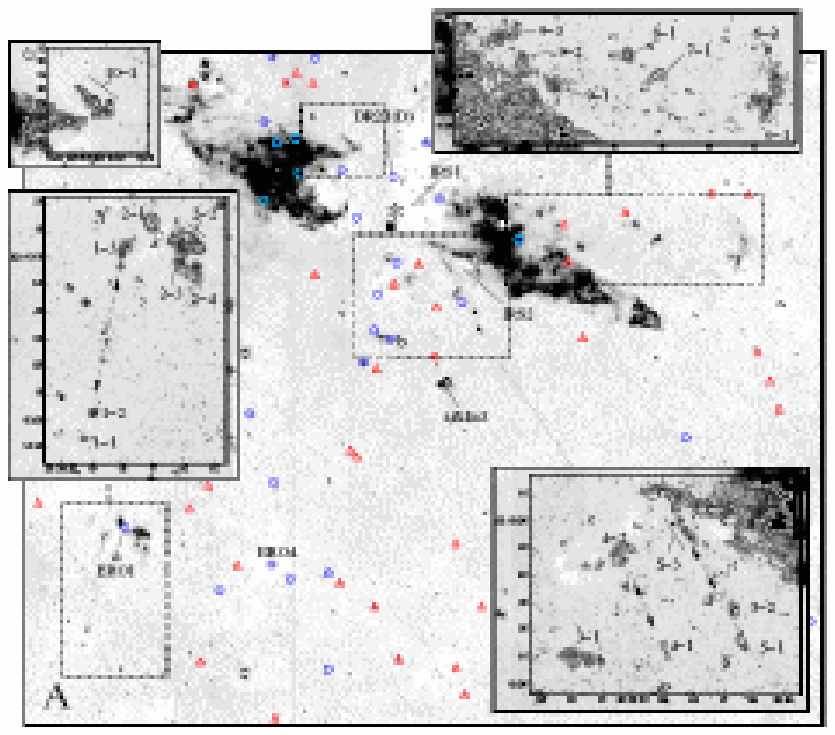}
 \vspace*{0.3cm} 
 \caption[] 
 {Continuum-subtracted H$_2$ 1-0S(1) image
 of region ``A'', just to the south of DR21.  Inset are contour-plots
 showing the prominent line-emission features in more detail.  The
 micro-stepped data have been binned over 2x2 pixels to a scale of
 0.4\arcsec\ to improve the signal-to-noise on the extended H$_2$
 features.  Contour levels measure 0.4, 0.8, 1.6$\times 10^{-18}$\Wma\
 (black) and 3.2, 6.4, 12.8$\times 10^{-18}$\Wma\ (white).  The open
 circles mark the locations of the protostars ($\alpha > 0.25$);
 the triangles mark the positions of the pre-main-sequence objects ($0.25 > \alpha
 > -1.60$).  Those that are possible jet sources are given reference
 letters and are also marked in the contour plots. The filled squares mark
 the positions of methanol masers in the DR21 outflow
 \citep{pla90,kog98}; the star indicates the location of the DR21
 A-B-C H{\sc ii} region (DR21 D is detected by Spitzer and is labeled).}  
 \label{h2jetsa} 
 \end{figure*}

\begin{figure*}
 \vspace*{0.1cm}
     \epsfxsize=17.5cm
     \epsfbox{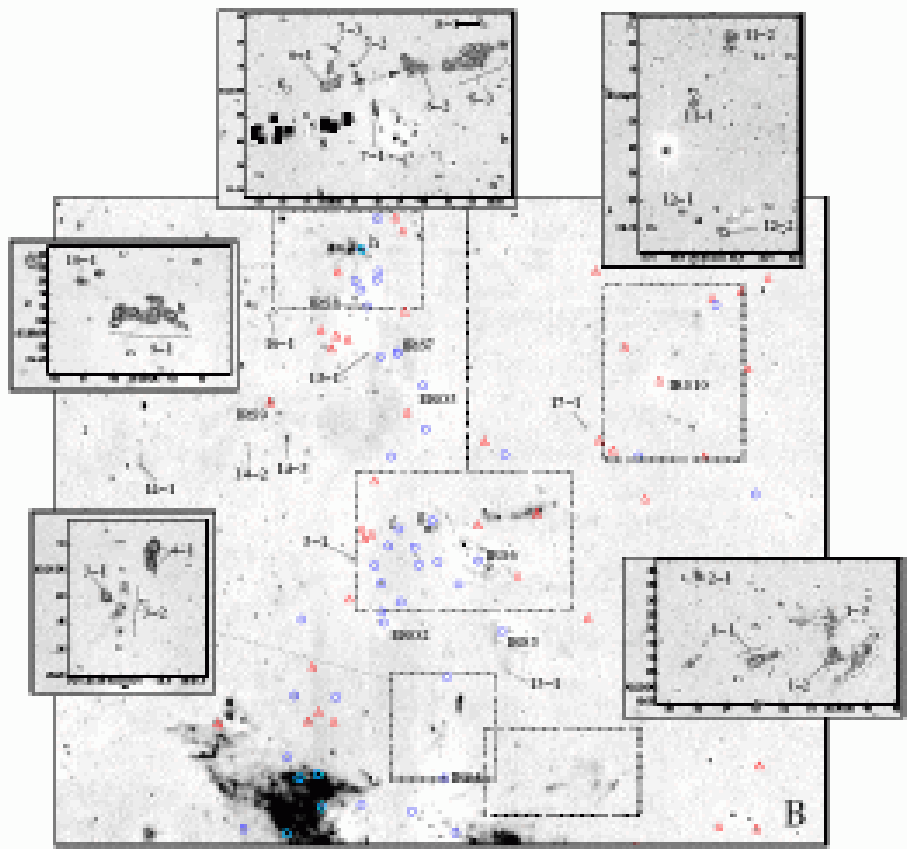}
 \vspace*{0.3cm} 
 \caption[] {Continuum-subtracted H$_2$ 1-0S(1) image
 of region ``B'' just to the north of DR21. The positions of
 YSOs are again marked with open circles and triangles; possible jet sources 
 are given letters. In the uppermost contour plot, methanol masers associated 
 with the DR21(OH) outflow are marked with filled boxes \citep{pla90,kog98}, while
 water masers are marked with $\times$ symbols \citep{for78,man92}. 
 (See Fig.\ref{h2jetsa} for further details.)} 
 \label{h2jetsb} 
 \end{figure*}

\begin{figure*} 
  \epsfxsize=16.0cm
  \epsfbox{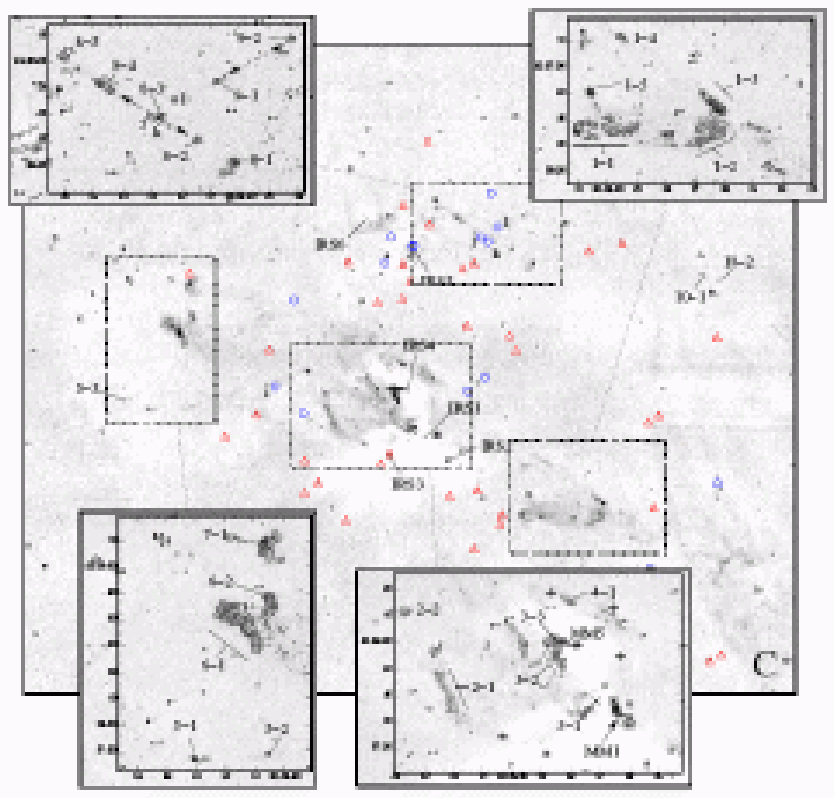}
 \vspace*{0.3cm} 
 \caption[] 
 {Continuum-subtracted H$_2$ 1-0S(1) image of region ``C'' centred on
 W75N (see Fig.\ref{h2jetsa} for additional details). YSOs are again
 marked with open circles and triangles; possible jet sources are
 assigned letters. In the bottom-right contour plot the
 positions of water masers are marked with $\times$ symbols \citep{hun94,tor97};
 2.7~mm continuum sources \citep{she03} are marked
 with pluses. Two of these mm sources, MM~1 and MM~5, coincide with
 W75N~IRS1/W75N(B) and W75N~IRS4/W75N(A), respectively.  }
 \label{h2jetsc} 
 \end{figure*}

\begin{figure*}
     \epsfxsize=16.5cm
     \epsfbox{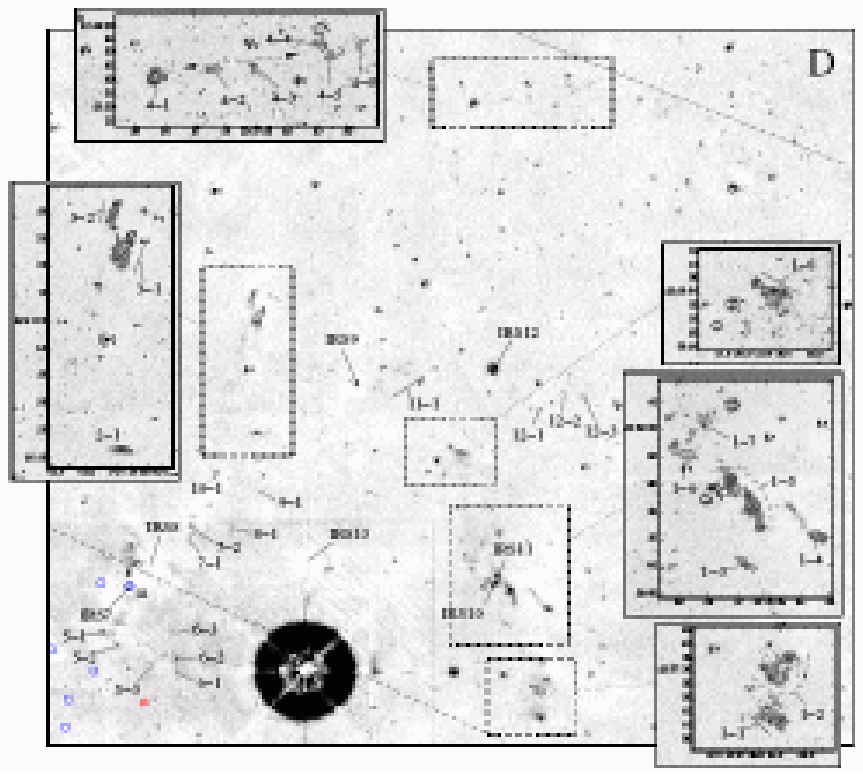}
 \vspace*{0.3cm} 
 \caption[] {Continuum-subtracted H$_2$ 1-0S(1) image of region ``D''
 $\sim$10\arcmin\ to the west-north-west of W75N. YSO positions are
 again marked with circles and triangles, although the Spitzer data
 are incomplete in this region; the dashed line marks the edge of the
 3.6\mic\ and 5.8\mic\ maps, while the dot-dashed line marks the edge of
 the 4.5\mic\ and 8.0\mic\ mosaics. (See Fig.\ref{h2jetsa} for further
 details.)}
 \label{h2jetsd} 
 \end{figure*}

\begin{figure*}
 \hspace*{0.3cm}
     \epsfxsize=15.5cm
     \epsfbox{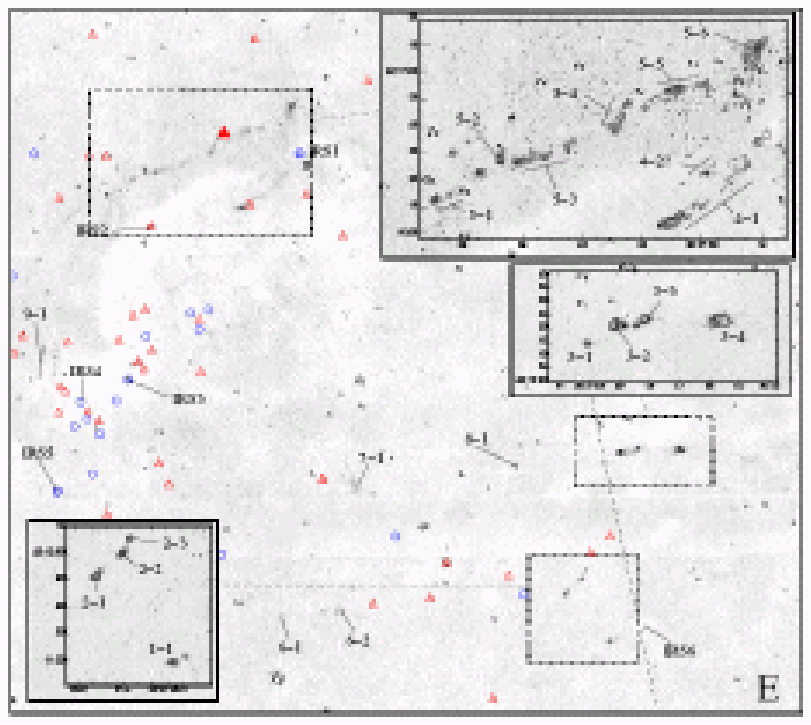}
 \vspace*{0.3cm} 
 \caption[] {Continuum-subtracted H$_2$ 1-0S(1) image of L906E, region
 ``E'' (which is $\sim$20\arcmin\ to the west of DR21).  The positions
 of Spitzer YSO candidates are marked with circles and triangles; those that may
 drive the H$_2$ jets are given letters.  The dashed line marks the 
 approximate edge of the
 four Spitzer mosaics. (See Fig.\ref{h2jetsa} for
 additional details.) The white thumb-print-like patch that coincides
 with L906E-IRS sources 1, 2 and 3 is an artifact from the subtracted
 K-band image.  It was produced by patches of dirt on the WFCAM field
 lens \citep{dye06}. }
 \label{h2jetse}
  \end{figure*}

In Figs.~\ref{h2jetsa} to \ref{h2jetse} we show continuum-subtracted
H$_2$ images of various regions along and adjacent to the DR21/W75
molecular ridge.  We have separated the data into five areas for
clarity, labeled A--E in Fig.~\ref{wfcam0}.  Individual H$_2$ knots,
filaments and bow-shocks within each area have been numbered.
Features that are thought to be part of the same outflow have been
given consecutive numbers.  For example, knots A~1-1, A~1-2 and A~1-3
are possibly part of the same jet in area A, while A~2-1, although
still in area A, is probably part of a different flow.  We also mark
on each figure the locations of Spitzer sources with reddened
photospheres; those with $\alpha > 0.25$ are dubbed protostars, and
those with $0.25 > \alpha > -1.65$ we refer to as pre-main-sequence
objects.  Note that only sources detected in all four Spitzer IRAC
bands were extracted from the data.


\subsection{DR21 south - region A}

The area just to the south of DR\,21 is shown in Fig.~\ref{h2jetsa}.
The region includes the 850\mic\ core and mid-IR peak ERO~1, and the
possible edge-on disk system ERO~4 discussed earlier.  ERO~1 (star
``a'' in Table~\ref{sources}) is identified as a reddened protostar
in our analysis of the Spitzer IRAC data (it is marked with an
open circle). Bright H$_2$ emission is associated with ERO~1, though
not with ERO~4.  Knots A~2-2, A~2-3 and A~2-4 outline a roughly
conical structure with the apex facing ERO~1.  This morphology could
be indicative of a wide flow opening angle.  The counter-flow is
traced by knot A~2-1. The flow axis is clearly orientated roughly
northeast-southwest.

A number of collimated H$_2$ jets are found scattered around the
region.  The jet described by A~1-1, A~1-2 and A~1-3 is almost
certainly unrelated to ERO~1. This jet has no Spitzer-detected central
engine (no open circle or triangle in Fig.~\ref{h2jetsa}).  Nor is it
associated with an 850\mic\ continuum peak.

Jet A~5-1 to A~5-3 and jet A~10-1 both point back roughly towards the
infrared cluster associated with DR21 (DR21-IRS~1 in Fig.~\ref{wfcam1}
and Fig.~\ref{h2jetsa}), and could therefore be driven by sources within
this region.  Alternatively, A~10-1 could be associated with the
luminous, accreting infrared source and H{\sc ii} region DR21~D
\citep{hsm05}, while knots A~5-1 and A~5-2 could be driven by the
bright K-band source DR21-IRS~2 (Fig.\ref{wfcam1}) or
the reddened YSO ``d'' found midway between knots A~5-1
and A~5-3.

A~4-1 and A~4-2 possibly represent a faint jet and bow shock, with YSO
``c'' as a tentative source candidate.  Notably, A~4-1:A~4-2 is
parallel with A~5-1:A~5-3.

A~9-1, A~9-2 and A~9-3 may delineate another collimated flow that is
roughly parallel with the main DR21 outflow, while knot A~6-1 may be a
bow shock driven by a source to the southeast.  Protostellar outflow
source candidates are not found in the Spitzer data, however.

A~3-1 is a fairly bright though non-descript feature in the centre of
the field.  It is close to the Spitzer YSO star ``b'', which is a
protostar with a notably steep spectral index
(Table~\ref{sources}).

Additional faint knots and features are identified around the region which
may or may not be associated with collimated flows.  In many cases,
the driving source of each flow is likely found along the molecular
ridge that runs north-south through DR21 (Fig.~\ref{scu}).

\subsection{DR21 north - region B}

Area B (Fig.~\ref{h2jetsb}) is centred on DR21(OH) and the
jet labeled B~6-1 to B~6-3.  Orthogonal to this flow we detect a
second, curving flow, knots B~7-1 to B~7-3.  Neither flow seems to be
associated with DR21(OH) itself: \citet{lai03} and \citet{val06} both
detect high-velocity CO emission associated with DR21(OH).  However,
the flow -- mapped with a resolution of a few arcseconds by
\citet{lai03} -- extends roughly east-west.  Knot B~5-1 and possibly
B~7-1 could be powered by this outflow.

Jet B~7-1 to B~7-3 could instead be driven by the bright, nebulous
near-IR source $\sim$40\arcsec\ to the west of DR21(OH), labeled
DR21-IRS~6 in Fig.\ref{wfcam1}. This source has Spitzer colours
consistent with it being a reddened protostar, and is given reference
letter ``e'' in Table~\ref{sources}.  Sources ``f'' or ``g'' may
likewise be associated with jet B~6-1:B~6-3, although the
pre-main-sequence YSO $\sim$10\arcsec\ east of B~6-2 is also well
placed. Stars ``f'' and ``g'' have particularly steep spectral
indices.  Comparison of our images with the NH$_3$ thermal emission
maps of \citet{man92} shows that these two jets lie roughly midway
between (and to the west of) two compact clusters of cores (labeled
MM~1/MM~2 and DR21(OH)N by this group).  There is also no obvious dust
clump associated with sources ``f'', ``g'' or DR21-IRS~6/source ``e'',
or with either jet, in the 850\mic\ map in Fig.~\ref{scu}.

To the south in this figure we identify at least two (though possibly
many more) collimated jets radiating outward from the DR21 core and IR
cluster: feature B~1-1 is probably a jet driving the knotty bow
B~1-2:B~1-3, while knot B~4-1 is a possible bow heading northward from
DR21-IRS~4 or a source in the vicinity of DR21 itself.  The compact
knot B~3-1 and the fainter features labeled B~3-2 are enveloped in
faint, diffuse emission.  These may delineate further collimated
flows extending northward from DR21.  Including the jets identified in
the previous section, we find {\em at least} five collimated jets radiating
outward from the DR21 850\mic\ peak and cluster of H{\sc ii} regions and
embedded IR sources (six outflows if we include the main DR21 outflow
itself).

Continuing northward along the molecular ridge, we find a chain of
bright H$_2$ knots, B~9-1, which may be part of an east-west flow
driven by the Spitzer YSO candidate source ``h'' at the western end of
the B~9-1 chain. Alternatively, this feature may represent a
fragmented bow shock in a flow driven by an embedded source to the
south.  Protostellar sources ERO~3, DR21-IRS~7 and DR21-IRS~8 are all
possible candidates.

There are also faint knots $\sim$20\arcsec\ south-east and
$\sim$20\arcsec\ north-east of DR21-IRS~10.  Although this star is
identified as a possible Class II or flat spectrum source in our Spitzer
analysis (Table~\ref{irs}), it is also associated with a discrete source
at 850\mic\ (Fig.~\ref{scu}).  Multiple unresolved sources may explain
these seemingly contradictory data.  IRS~10 certainly seems to be a
strong source candidate for the collimated flow associated with B~11-1
and B~11-2.

There are other faint H$_2$ features distributed east of the
ERO~3/FIR~1, FIR~2 and FIR~3 chain of dust cores, which we identify in
the main panel in Fig.~\ref{h2jetsb}.  These are also evident in
Fig.~\ref{wfcam1}.  On morphological grounds alone, associating them
with embedded sources along the molecular ridge is very difficult.
Suffice to say that these features are generally within an arcminute
($\sim$0.9~pc) of the molecular ridge and may be part of east-west
orientated jets.

\subsection{W75N - region C}

In Fig.~\ref{h2jetsc} we show a continuum-subtracted H$_2$ image of
W75N.  The curving filaments of H$_2$ emission associated with the
ends of the large-scale bipolar CO outflow, first identified by
\citet{dav98a}, are once again evident (features C~1-1 to C~1-4 
and C~6-1:C~6-2).  \citet{tor97} suggest that this large-scale
flow could be driven by VLA\,1, based on the elongated structure of
this source at radio wavelengths.  The source is well aligned with the
apices of the proposed bow shocks C~1-1 to C~1-4 and C~6-1 and C~6-2.
Alternatively, knot C~1-5, within the wings of the C~1-1 to C~1-4 bow,
may be part of a collimated jet driving the bow shock, driven by the
reddened K-band source W75N-IRS~2 (Fig.~\ref{wfcam2}).  Hydrogen
recombination line emission has been detected from this star, so the
source may be undergoing mass loss \citep{moo91b}.  (The faint, diffuse
emission a further arcminute to the south-west of knot C~1-2, noted by
\citet{she03}, is also evident in the main panel in Fig.~\ref{h2jetsc}, 
and as faint red features in Fig.~\ref{wfcam2}.)

Knot C~7-1 to the north of the main W75N counterflow lies along the
axis of the extended knot C~4-1 and may be part of the same flow.  A
Spitzer YSO candidate and three dust cores traced in 2.7~mm continuum
emission \citep{she03} all lie on approximately the same axis
(although the millimetre peaks have no bright K-band nor 8.0\mic\
counterparts).  There are regions of diffuse H$_2$ emission between
C~4-1 and C~7-1 (labeled NE-H and NE-I by \citet{she03}) which may
also be part of this flow.  Alternatively, this diffuse emission may
be non-thermal, i.e. ambient gas fluorescently excited by neighbouring
B stars.

A further $\sim$2\arcmin\ to the north, knots C~8-1 to C~8-5 trace a
very well collimated jet with almost the same position angle on the
sky as the main W75N molecular outflow and the tentative C~4-1 to
C~7-1 jet.  However, if features C~9-1 and C~9-2 are also part of a
collimated jet, this flow is almost orthogonal to the other outflows.
Two Spitzer YSO sources ``j'' and ``k'' are candidates for
C~8-1:C~8-5, while source ``l'' likely powers C~9-1:C~9-2.  In our
K-band image ``k'' and ``l'' are both slightly extended, their conical
nebulae opening in the directions of the C~8-1:C~8-5 and C~9-1:C~9-2
jets respectively.  All three protostars have steep spectral indices
(Table~\ref{sources}), and together these three stars are associated
with a compact molecular core, labeled W75N-N in the 850\mic\ map in
Fig.~\ref{scu} (see also Table~\ref{scuba}).

We also mention knots C~5-1 and C~5-2 on the southern edge of the main
W75N outflow.  These features form part of a very faint, narrow
structure that snakes back toward W75N-IRS~3.  However, one could also
argue that C~5-1 and C~2-2 form part of a collimated jet, powered by
source ``i'', which notably coincides with SCUBA core W75N-E
(Table~\ref{scuba}; although this peak may actually trace CO line
emission in SCUBA 850\mic\ filter bandpass, rather than dust continuum
emission).

Subtraction of the diffuse continuum that envelopes W75N(A) and (B)
reveals a number of elongated H$_2$ structures.  These are thought to
be bona fide line-emission features, since they are also recovered in
the H$_2$ images of \citet{dav98a} and \citet{she03}.  The collimated
structure C~2-1 and knot C~2-2 may be part of another collimated jet,
although there is no bright K-band nor 8.0\mic\ source along this
axis. Knots C~3-1 and C~3-3 may be bow shocks in the lobes of a
bipolar jet, seen here as knotty feature C~3-2.  Such a jet might be
driven by a source midway between C~3-1 and C~3-3, although again no
millimetre nor 8.0\mic\ peak is found here. 

Shepherd and collaborators devote a series of papers to examining the
properties of the sources in the central W75N cluster, and to
identifying the source or sources that drive the large-scale CO outflow
\citep{she01,she03,she04}.  They identify a possible complex of multiple, 
overlapping, small-scale outflows in W75N.  They also point out that
the VLA~1 jet lacks sufficient momentum to drive the large-scale CO
outflow, and conclude that multiple sources may be producing the flow.
Although our new near-IR images support the idea of multiple flows in
W75N (and its immediate surroundings), the data do not really shed any
further light on this issue.

\subsection{W75N west - region D}

Area D, to the west of W75N, was not observed in all four IRAC bands.
Therefore, we can not use IRAC photometry to search for embedded YSOs.
Even so, the region is of considerable interest.  It includes a
prominent bipolar molecular outflow labeled D~1-1 to D~1-8.  On
morphological grounds alone it seems likely that features D~1-1:D~1-2
and D~1-8 represent bow shocks at the ends of bipolar flow lobes.
Midway between these features there is a compact cluster of at least
three near-IR sources.  Two targets, W75N-IRS~10 and 11, are also
8.0\mic\ point sources and are strong candidates for the outflow
central engine (see also Fig.~\ref{spit3}). W75N-IRS~11 in particular
has a [4.5]-[8.0] colour of $\sim$3.1 (Table~\ref{irs}) and is
coincident with H$_2$ emission.

The elongated H$_2$ ``fingers'' D~3-1 and D~3-2 in Fig.~\ref{h2jetsd}
are reminiscent of the ``bullets'' in Orion (and perhaps knot B~4-1).
These may originate from the bright K-band/8.0\mic\ sources
W75N-IRS~7 (source ``m'') or W75N-IRS~8.  The latter was only observed
in two IRAC bands, though like W75N-IRS~7 it is quite red
(Table~\ref{irs}).

Further to the north, the H$_2$ features D~4-1 to D~4-6 again describe
a collimated jet.  Features D~4-4 and D~4-5 are probably the wings of
-- and D~4-6 the cap of -- a westward-moving bow shock.

In Fig.~\ref{wfcam2} and \ref{spit3} we label four other interesting
K-band/8.0\mic\ sources, W75N-IRS~9, 12, 13 and 14.  There is very
faint H$_2$ emission between sources W75N-IRS~9 and 12, labeled
D~11-1 in Fig.~\ref{h2jetsd}.  Additional faint H$_2$ jets and knots
are detected west of W75N-IRS~12 and around W75N-IRS~7/IRS~8. Sources
13 and 14 appear to be point sources within small ionised shells which
we see clearly in Fig.~\ref{wfcam2} in scattered light and/or
recombination-line emission.  There is possibly some fluorescent H$_2$
emission along the edges of each shell.  We see no evidence for H$_2$
jets associated with either source, however.

\subsection{L906E - region E}

Finally, to the west of DR21 there is a cluster of nebulous embedded
sources and a bok-globule-like region of increased extinction in the
near-IR.  These features lie on the eastern edge of the Lynds dark
cloud L906.  Unfortunately this area is outside the field of view of
the SCUBA 850\mic\ map, and it was not discussed in great detail by
\citet{mar04}.  However, in Fig.~\ref{spit4} and Fig.~\ref{h2jetse} 
one easily sees the rich nature of this region.

We clearly identify a number of collimated jets in area E. The
wishbone-shaped feature E~4-1, E~4-2 and E~5-1 to 5-6 is associated
with a bright 8\mic\ peak and spherical core of mid-IR emission, which
we dub L906E-IRS~1.  This source has a particularly steep spectral
index and certainly qualifies as an ERO (Table~\ref{sources}). Even
so, the H$_2$ knots are probably associated with independent flows,
rather than a single wide-angled wind.

The H$_2$ features E~3-1 to E~3-4 are unfortunately beyond the edges
of all four of the Spitzer IRAC images, so these cannot be used to search for
the possible outflow source.  However, they likely form part of a knotty
jet.  So might E~2-1 to E~2-3, which lie 20\arcsec --40\arcsec\ to the
southeast of a pair of Class II/flat spectrum sources.

The faint knots E~6-1 and E~6-2 likely trace yet another collimated
flow, while feature E~7-1 appears to be a faint bow shock, possible
driven by a source in the vicinity of L906E-IRS~3.

Of the remaining H$_2$ feature, E~1-1 is close to a nebulous K-band
source L906E-IRS~6 (situated $\sim$25\arcmin\ to the west-northwest of
E~1-1).  This source was just inside the 5.8\mic\ and 8.0\mic\
Spitzer field of view, and again appears to be fairly red
(Table~\ref{irs}).

\bsp

\label{lastpage}

\end{document}